\documentclass[11pt]{article}
\pdfoutput=1

\usepackage[no-natbib-sort]{jheppub}
\usepackage[T1]{fontenc} 

\usepackage{amssymb}
\usepackage{amsfonts}
\usepackage{amsbsy}
\usepackage{amsmath}
\usepackage{amsthm}
\usepackage{graphicx}
\usepackage{epstopdf}
\usepackage[vcentermath]{youngtab}
\usepackage{multirow}
\usepackage{latexsym} 
\usepackage{array}
\usepackage{color}
\usepackage{verbatim}
\usepackage[utf8]{inputenc}

\input cyracc.def %sha
\newfont{\twelvecyr}{wncyr10 at 12pt}
\def\Z{\mathbb{Z}}
\def\F{\mathbb{F}}

\def\C{\mathbb{C}}

\def\P{\mathbb{P}}

\def\n3a{t}

\def\mred{M_{\rm red}}

\newcommand{\SU}[0]{\mathrm{SU}}
\newcommand{\SO}[0]{\mathrm{SO}}
\newcommand{\U}[0]{\mathrm{U}}

\newcommand*{\cy}{CY }
    \newcommand*{\gsm}{G_{\rm SM}}

\newcommand{\drop}[1]{\footnote{{\color{red} dropped material:  #1}}}

\title{Gauge symmetry breaking with fluxes and
 natural Standard Model structure from exceptional GUTs in F-theory}

\author{Shing Yan Li}
\author{and Washington Taylor}

\affiliation{Center for Theoretical Physics\\
Department of Physics\\
Massachusetts Institute of Technology\\
77 Massachusetts Avenue\\
Cambridge, MA 02139, USA}

\emailAdd{sykobeli at mit.edu}
\emailAdd{wati at mit.edu}

\preprint{MIT-CTP/5407}
\abstract{
We give a general description of gauge symmetry breaking using
vertical and remainder fluxes in 4D F-theory models.
The fluxes can break a geometric gauge group to a smaller group and induce chiral
matter, even when the larger group admits no chiral matter
representations.  We focus specifically on applications to
realizations of the Standard Model gauge group and chiral
matter spectrum through breaking of rigid
exceptional gauge groups $E_7, E_6$, which are ubiquitous in
the 4D F-theory landscape. Supplemented by an intermediate
$\SU(5)$ group, these large classes of models give
natural constructions of Standard Model-like theories with small
numbers of generations of matter in F-theory.
}
  
\begin{document}
\maketitle
\flushbottom

%--------------------------------
\section{Introduction}
\label{sec:Intro}

String theory provides a consistent framework for a unified theory
that combines gravity with the other fundamental forces described by
quantum field theory.  To describe the real world, however,
ten-dimensional string theory must be compactified on a real
six-dimensional manifold, and various further objects like branes,
flux, and orientifolds must be incorporated.  Such constructions give
an enormous number (perhaps on the order of something like
$10^{272000}$ \cite{TaylorWangVacua}) of string theory vacua, known as
the string landscape.  Despite this large number, so far it has not
been clear which low-energy theories can be UV-completed and realized
in the string landscape.  The investigation of this question, known as
the Swampland program \cite{VafaSwamp,OoguriVafaSwamp}, has been a
rapidly evolving research area.

Here we focus on another related main challenge in string
phenomenology.  Despite decades of work, it is not yet clear whether
the well-established Standard Model of particle physics (SM) can be
realized in the string landscape, including all details of observed
phenomenology; for a recent review of work in this direction, see
\cite{Cvetic:2022fnv}.  Beyond the simple existence of such a solution, it is
perhaps even more important to understand the extent to which the
Standard Model can arise as a \emph{natural} solution in string
theory.  In other words, we would like to understand the extent to
which solutions like the Standard Model are widespread in the string
landscape or require extensive fine-tuning.  Constructing the detailed
Standard Model requires many elements such as the gauge group, the
matter content including both chiral matter and the Higgs, the Yukawa
couplings, a supersymmetry (SUSY)-breaking mechanism, values of the 19
free parameters, and possibly some room to address beyond-SM problems
as well as cosmological aspects such as the density of dark energy.
Unfortunately, the current available string theory techniques are far
from enough to compute all these features precisely. Although there is
some recent development on finding the exact matter spectrum
\cite{Bies:2014sra,Bies:2021nje,Bies:2021xfh,Bies:2022wvj}
in F-theory, in this paper we only
focus on the gauge group and chiral matter content, where the
techniques have been well developed.  The general philosophy is that
if we can identify a natural class of models that realize the Standard
Model gauge group and chiral matter fields, these structures may
naturally correlate with certain other features of SM or beyond SM
physics.

These aspects have been long-standing and primary goals in string
phenomenology, and there has been a great amount of work on them in
the last two decades, starting from heterotic string compactifications,
which naturally carry $E_8$ gauge groups that can be broken down to
the standard model gauge group.  Recently, F-theory
\cite{VafaF-theory,MorrisonVafaI,MorrisonVafaII} has become the most
promising framework for studying string compactifications and
phenomenology, as it provides a global description of a large
connected class of supersymmetric string vacua. (See
\cite{WeigandTASI} for a review.) In particular, F-theory gives 4D
$\mathcal N=1$ supergravity models when compactified on 
elliptically fibered Calabi-Yau (CY) fourfolds, corresponding to
non-perturbative compactifications of type IIB string theory on general
(non-Ricci flat) complex K\"ahler threefold base manifolds $B$.  The
number of such threefold geometries $B$ alone seems to be on the order
of $10^{3000}$
\cite{TaylorWangMC,HalversonLongSungAlg,TaylorWangLandscape}, without
even considering the exponential multiplicity of fluxes possible for
each geometry.  F-theory is also known to be dual to many other types
of string compactifications such as heterotic models.  Briefly,
F-theory is a strongly coupled version of type IIB string theory with
non-perturbative configurations of 7-branes balancing the curvature of
the compactification space. The non-perturbative brane physics is
encoded geometrically into the elliptically fibered manifold, which
can be analyzed using powerful tools from algebraic geometry. The
gauge groups and matter content supported on these branes can then be
easily determined when combined with flux data. Applying these
techniques, here we construct a novel class of F-theory models that
naturally give the SM gauge group and chiral matter content.
Note that in this paper we focus exclusively on 4D models with ${\cal
  N} = 1$ supersymmetry (4 supercharges).  While  low-scale supersymmetry has not been
observed in nature, supersymmetry provides additional symmetry
structure that enables systematic analytic study of a broad class of
vacua; since some structure such as typical rigid gauge groups are
similar between 6D theories with 8 supercharges and 4D theories with 4
supercharges, we have some optimism that some of the structure of
typical geometric gauge groups and chiral matter content may persist
from 4D  ${\cal N} = 1$ theories to theories with broken supersymmetry.

There have been many attempts in the literature to construct
supersymmetric
models of compactified string theory
with the SM gauge group $\gsm=\SU(3)\times \SU(2)\times \U(1)/\mathbb
Z_6$.
(As noted in \cite{TaylorTurnerGeneric,TaylorTurner321}, for the gauge
group $\SU(3) \times\SU(2) \times\U(1)$
without the quotient by the $\Z_6$ center, the SM chiral
matter content is highly non-generic and involves a great deal of fine
tuning; we proceed under the assumption that the gauge group of
the Standard Model is really $\gsm$.)
The results of these efforts suggest that the (supersymmetric)
landscape may contain a wide variety of SM-like models. The constructions of such
models in F-theory can be loosely classified in the following ways: As in field
theory approaches, one can directly build models with $\gsm$, or start
with grand unified theories (GUTs) and break the larger gauge group
down to $\gsm$ in various ways. There are also two essentially
distinct types of geometric gauge groups in F-theory. On the one hand,
one can tune a desired gauge group by fine-tuning many complex
structure moduli. In contrast, most F-theory compactification bases
contain divisors with very negative normal bundle. The strong
curvature of these geometries forces singularities in the elliptic
curve over these divisors, giving rigid (a.k.a \emph{geometrically
  non-Higgsable} \cite{MorrisonTaylor4DClusters}) gauge symmetries,
which are present throughout the whole branch of moduli space and
ubiquitous in the F-theory landscape
\cite{TaylorWangMC,HalversonLongSungAlg,TaylorWangLandscape}. Below we
comment on each type of approach:

\begin{itemize}
    \item \textbf{Directly tuned $\gsm$:}
      These models do not require any symmetry
      breaking mechanisms except the usual Higgs. Recently significant
      progress on these has been gained. In
      \cite{CveticEtAlQuadrillion}, $10^{15}$ explicit solutions of
      directly tuned $\gsm$ with three generations of SM chiral
      matter (a ``quadrillion Standard Models''), have been
      constructed, based on the ``$F_{11}$'' fiber of
      \cite{KleversEtAlToric};. It has also been shown that the SM
      matter representations generically appear when $\gsm$ is
      directly tuned, in the sense that these matter representations
      are included among those that require the least amount of moduli
      fine-tuning given the gauge group \cite{TaylorTurnerGeneric},
      and a universal Weierstrass model for such tunings has been
      constructed \cite{Raghuram:2019efb}, which includes those of
      \cite{CveticEtAlQuadrillion} in one particular subclass.  All
      these constructions include the presence of the
 $\mathbb Z_6$
      quotient in $\gsm$.
    
    \item \textbf{Directly tuned GUT:}
    These models have been studied for over a decade, starting
    from
    \cite{Donagi:2008ca,BeasleyHeckmanVafaI,BeasleyHeckmanVafaII,DonagiWijnholtGUTs}. Most
    of the work on these models has focused on the GUT
    group of $\SU(5)$ and its $\U(1)$ extensions
    \cite{Blumenhagen:2009yv,Marsano:2009wr,Grimm:2009yu,KRAUSE20121,Braun:2013nqa},
    while there has also been some study of
    $\SO(10)$ GUTs \cite{Chen:2010ts}. (See \cite{HeckmanReview} for a
    review) Most of these constructions break the GUT group using the
    so-called hypercharge flux further discussed in
    \cite{Mayrhofer:2013ara,Braun:2014pva}, which is a kind of ``remainder''
    flux \cite{Braun:2014xka} breaking the gauge group into the commutant of broken
    directions, including the $\U(1)$'s of these directions \cite{Buican:2006sn}.
    
    \item \textbf{Rigid $\gsm$:}
    Despite the success of the above models, they cannot be
    the most generic or natural SMs in the landscape, as
    extensive fine-tuning is generally required to get the directly
    tuned $\gsm$ or a non-rigid (tuned)
GUT group such as  $\SU(5)$ (see
    e.g.\ \cite{BraunWatariGenerations}).
    Moreover, the presence of rigid gauge groups forbids
    tuning additional gauge factors like $\gsm$ on most bases.
    Finding a rigid $\gsm$ seems to be a more natural way.
    Nevertheless, while the non-abelian $\SU(3)\times \SU(2)$
    parts of $\gsm$ can easily
be realized as a rigid structure \cite{GrassiHalversonShanesonTaylor},
    constructing the $\U(1)$ is much more subtle, and bases that support
    non-Higgsable $\U(1)$ factors are rather rare
    \cite{MartiniTaylorSemitoric, WangU1s}.
    
    \item \textbf{Rigid GUT:} The rigid gauge groups that contain
      $\gsm$ as a subgroup are $E_8,E_7,E_6$
      \cite{MorrisonTaylor4DClusters}, and these rigid groups are
      ubiquitous in the F-theory landscape.  Of these, it seems that
      in 4D (as well as in 6D),
      $E_8$ appears most frequently in the landscape, while $E_7$ and
      $E_6$ are also quite abundant
      \cite{TaylorWangMC,HalversonLongSungAlg,TaylorWangLandscape}. Starting
      with one of these rigid exceptional groups and breaking down to
      $\gsm$ is in principle the most natural way to construct SM-like
      models, from the point of view of prevalence in the F-theory
      landscape, and this is the approach taken in this paper. On the other hand,
      SM-like models using these groups bring other
      challenges. Undesired exotic matter can be easily induced by
      such large gauge groups. While $E_6$ has been one of the
      traditional GUT groups 
(see, e.g., \cite{Gursey:1975ki,Achiman:1978vg,Barbieri:1980vc}, and
\cite{Chen:2010tg,Callaghan:2012rv,Callaghan:2013kaa}
for realizations in F-theory and further references)
, $E_7$ and $E_8$ do not themselves
      support chiral matter and have not received as much attention as
      GUT groups.  These groups, especially $E_8$,
are often associated with high
      degrees of singularity in the elliptic fibration (i.e.,
      codimension two  (4, 6) loci), that involve
      strongly coupled sectors that are poorly
      understood \cite{HeckmanMorrisonVafa,Apruzzi:2018oge}; the
      constructions we consider here avoid these issues.
    
\end{itemize}

Recently in \cite{Li:2021eyn}, we have proposed a general class of
SM-like models using a rigid (or even tuned) $E_7$ GUT group in
F-theory, with an intermediate $\SU(5)$ group. These models enjoy the advantages of being natural and
requiring little fine-tuning, and address some of the above
challenges.  Specifically, fluxes can be used to break the
geometric $E_7$ group in an F-theory construction in a way that is not
transparent in the low-energy field theory, but
%we can go beyond the field theory limit and
%use $E_7$ as GUT group, 
gives the correct SM gauge group and some chiral matter.
Although in many cases the breaking leads to exotic chiral
matter, there are large families of models in which the
correct SM chiral matter representations are obtained through
an intermediate $\SU(5)$.  The number of generations
can easily be small and we have demonstrated that three
generations can naturally arise in many of these models.  In this longer followup, we present
the general formalism and various technical subtleties, describe the
$E_7$ models in much more detail, and generalize the construction to
other groups such as $E_6$. In particular, we give a fully
explicit example of our SM-like models, incorporating both
vertical and remainder fluxes.  These constructions open large new
regions of the landscape for string phenomenology. Note that for
various reasons explained below, we do not include $E_8$ GUTs,
although it is the most frequent exceptional gauge factor in the landscape.

The central tool we use to construct these models is gauge
symmetry breaking by flux living in \emph{vertical} and \emph{remainder} cohomologies (we
use the name ``flux breaking'' from now on; vertical and remainder cohomologies are reviewed in
\S\ref{subsec:flux}). This is an economic
way to deal with some of the above challenges. 
By imposing
simple linear constraints, we can break the larger GUT group
down to $\gsm$ without extra $\U(1)$'s. At the same time, the \emph{vertical} flux
induces chiral matter regardless of whether the original group
supports chiral matter.
%As we will show below, there are many
%gauge-breaking flux parameters contributing to a single chiral
%index.
The resulting chiral index has a linear Diophantine
structure related to the geometry of the F-theory base that generically allows any small number of
generations; sometimes three is the most
preferred number of generations. Remarkably, no highly tuned geometry
or nontrivial quantization condition on
the manifold is needed to achieve
 this structure. Certainly the idea of vertical
flux breaking is not new, but below we develop it to some
depth so that only a relatively simple calculation is needed to find the
chiral index. The calculation is based on the techniques in
\cite{Jefferson:2021bid}, which provide a conjecturally resolution-independent
description of the mathematical structure needed to compute chiral
indices for a fixed gauge group structure on a general base.

This paper is organized as follows: In
Section \ref{sec:Review}, we review the elements from F-theory that are
essential for constructing our SM-like models. We start by
describing the elliptic fibration of a general F-theory model
as a Weierstrass model. We write down the methods to determine
the gauge groups and matter representations from singularities
in the fibration. We also discuss the difference between tuned
and rigid gauge groups. Then we review the notion of vertical and remainder
fluxes, discuss various constraints for consistent flux
compactifications, and summarize how the framework
of intersection theory can be used as
the main tool to organize and
solve the flux constraints. 

After these preparations, we are ready to describe the formalism of
flux breaking in Section \ref{sec:Formalism}. There we write down the
flux constraints for gauge breaking and the formula for chiral
indices. We also describe various technical points such as determining
matter surfaces, primitivity and K\"ahler moduli stabilization. To
demonstrate how the formalism works, we work out several simple
$\SU(N)$ examples focusing on anomaly cancellation.

In Section \ref{sec:SM} we present the construction of natural SM-like models from $E_7$ flux breaking. These
models are also described in \cite{Li:2021eyn}, but we provide more details
here. We first discuss different embeddings of $\gsm$ into
$E_7$, which induce SM chiral matter or various exotic matter.
Then we write down the class of SM-like models in general,
without assuming a specific base.

The same method can be
straightforwardly generalized to other large gauge groups such as
$E_6$.  We discuss these applications in Section
\ref{sec:othergroups}. There we also discuss some obstructions to
applying the same formalism to $E_8$. As a useful example, in Section \ref{sec:example} we 
work out an explicit construction on a particular
base that can give three generations of SM chiral
matter as the minimal and preferred chiral spectrum.
This construction is the simplest example
that we are aware of
where all the ingredients in our class of SM-like models can be
realized. Note that as mentioned above,
these SM-like models are far from complete to really describe our
Universe. In Section \ref{sec:Conclusion} we finally conclude and
discuss further questions in these directions. We address several
technical points in Appendix \ref{sec:Stuckelberg}, \ref{sec:embeddingcount}, and
\ref{sec:resolution}.

\section{Review of F-theory}
\label{sec:Review}

In this section, we briefly review some general aspects of 4D
F-theory compactifications. These include the geometry of
elliptic fibrations and the associated $G_4$ flux. We only
discuss these issues to an extent that allows us to 
explain the construction of our class of
SM-like models. For more details of F-theory in general, we
refer readers to the excellent review by Weigand \cite{WeigandTASI}. 
The methods we use for working with
fluxes and chiral indices
follow the approach and notations of \cite{Jefferson:2021bid}.

\subsection{Basics of F-theory}
\label{subsec:Basics}

A 4D F-theory model
\cite{VafaF-theory,MorrisonVafaI,MorrisonVafaII} 
is associated with an elliptically fibered
CY fourfold $Y$ over a threefold base $B$. Such a model
 can be
considered as a non-perturbative type IIB string
compactification on $B$, where the shape of the elliptic fiber at each
point $x \in B$ is encoded by the IIB axio-dilaton $\tau (x) = \chi (x)
+ i e^{-\phi (x)}$. 
There is also a dual M-theory picture on the
resolved fourfold $\hat Y$;
 the 4D F-theory limit of 
the 3D
M-theory compactification on $\hat{Y}$ is taken when the elliptic
fiber shrinks to zero volume on the M-theory side. 
While much of the
physics of F-theory models is currently best understood using the
dual M-theory picture, the resolution of the geometry is not physical
in 4D,
and all this physics should in principle have a complete description
in the non-perturbative type IIB theory. 
Note that $B$ is in general a
compact K\"ahler manifold, but is not required to be CY. 
So the anticanonical class $-K_B$ need not vanish, but must be
effective for a good F-theory compactification to be possible.

The elliptic fibration in a general F-theory model can be
described by treating the elliptic curve parameterized by
$\tau (x)$ as a (1D) CY
hypersurface in the ambient projective space $\mathbb P^{2,3,1}$ with homogeneous coordinates $[x:y:z]$. The
fourfold $Y$ is then given by the locus of
\begin{equation}
    y^2=x^3+fxz^4+gz^6\,,
\end{equation}
where $f,g$ are sections of line bundles $\mathcal O(-4K_B),\mathcal O(-6K_B)$ respectively. This is known as a
Weierstrass model. The elliptic fiber
becomes singular when the discriminant
\begin{equation}
    \Delta=4f^3+27g^2\,,
\end{equation}
vanishes. In type IIB language, these vanishing loci represent
the positions of 7-branes, which are the sources for the singular
axio-dilaton background.

Consider a base divisor (algebraic subspace at codimension one in the base)
given by an irreducible codimension-one locus $\Sigma=\{s=0\}$ contained within
the vanishing locus of $\Delta$. The degree of the fiber singularity
at generic points on the divisor $\Sigma$ is determined by the orders
of vanishing of $f,g,\Delta$. When the orders are sufficiently high,
the fourfold $Y$ itself becomes singular, and a non-abelian gauge
group $G$ is supported on the divisor. We call such a divisor a gauge
divisor.  In general we abuse notation and use $\Sigma$ to denote both
the divisor and its homology class. The ``geometric'' gauge group, up to monodromies, can be
determined by the vanishing orders according to the classification by
Kodaira and N\'eron \cite{Kodaira,Neron,BershadskyEtAlSingularities}
(see Table \ref{Kodaira}). This geometry, however, does not fully
determine the physical gauge group since, as described below, it may
be broken by a flux background.  In this paper, we only consider
models with a single geometric
non-abelian gauge factor. The same kind of
analysis directly generalizes to the case of multiple geometric non-abelian
gauge factors, as the gauge divisors are just local features in the
geometry of $B$, although there can be further complications when geometric
non-abelian gauge factors intersect.
In principle, we expect that there may be a similar flux breaking story in
the presence of (Mordell-Weil) $\U(1)$ factors, although it may be
technically more involved and we leave exploration of such
constructions as a problem for the future.

\begin{table}[t]
    \centering
    \begin{tabular}{|c|c|c|c|c|c|}
    \hline
    Type & ord($f$) & ord($g$) & ord($\Delta$) & Singularity & Symmetry algebra \\
    \hline\hline
    $I_0$ & $\geq 0$ & $\geq 0$ & 0 & / & / \\
    $I_1$ & 0 & 0 & 1 & / & / \\
    $II$ & $\geq 1$ & 1 & 2 & / & / \\
    $III$ & 1 & $\geq 2$ & 3 & $A_1$ & $\mathfrak{su}(2)$ \\
    $IV$ & $\geq 2$ & 2 & 4 & $A_2$ & $\mathfrak{sp}(1)$ or $\mathfrak{su}(3)$ \\
    $I_n$ & 0 & 0 & $n\geq 2$ & $A_{n-1}$ & $\mathfrak{sp}([n/2])$ or $\mathfrak{su}(n)$ \\
    $I^*_0$ & $\geq 2$ & $\geq 3$ & 6 & $D_4$ & $\mathfrak{g}_2$ or $\mathfrak{so}(7)$ or $\mathfrak{so}(8)$ \\
    $I^*_n$ & 2 & 3 & $n\geq 7$ & $D_{n-2}$ & $\mathfrak{so}(2n-5)$ or $\mathfrak{so}(2n-4)$ \\
    $IV^*$ & $\geq 3$ & 4 & 8 & $E_6$ & $\mathfrak{f}_4$ or $\mathfrak{e}_6$ \\
    $III^*$ & 3 & $\geq 5$ & 9 & $E_7$ & $\mathfrak{e}_7$ \\
    $II^*$ & $\geq 4$ & 5 & 10 & $E_8$ & $\mathfrak{e}_8$ \\
    \hline
    non-min & $\geq 4$ & $\geq 6$ & $\geq 12$ & \multicolumn{2}{|c|}{incompatible with CY condition} \\
    \hline
    \end{tabular}
    \caption{Kodaira classification of singular elliptic fibers, mapping vanishing orders to non-abelian gauge groups up to monodromies.}
    \label{Kodaira}
\end{table}

As the geometry of $Y$ becomes singular in the presence of a
non-abelian gauge divisor $\Sigma$, to have well-defined geometric
quantities such as intersection numbers for the geometry, 
the usual procedure is to follow the M-theory approach and to
blow up the singular locus by $\mathbb P^1$'s, resulting in a smooth
resolved CY fourfold $\hat Y$. The resolution introduces a set of
exceptional divisors $D_i$ ($i=1,2,...,\mathrm{rank}(G)$) in the
fourfold, which are the $\mathbb P^1$-fibers over $\Sigma$. These new
divisors correspond to the Dynkin nodes of the group supported on
$\Sigma$, and their intersections match with the structure of the
Dynkin diagram. 
In accord with the Shioda-Tate-Wazir theorem
\cite{shioda1972,Wazir}, the divisors $D_I$ on $\hat Y$ are
spanned by the zero section\footnote{In general there are also
  divisors associated with abelian $\U(1)$ gauge factors when the
  fourfold $Y$ has a Mordell-Weil group of rational sections with nonzero
  rank.  In this paper we focus on geometries with only a single
  non-abelian gauge factor and no global $\U(1)$ factors from
  Mordell-Weil structure.} ($[x:y:z]=[1:1:0]$) $D_0$, the pullbacks
of base divisors $\pi^*D_\alpha$ (which we also call $D_\alpha$
depending on context), and exceptional divisors $D_i$. 
Notice that
there is no unique choice of the resolution and $D_i$'s, although 
consistency of the theory requires that the physics is independent of such a
choice. 
The resolution independence of the physics and of certain relevant
aspects of
the intersection form on CY fourfolds
(as found in \cite{Jefferson:2021bid} and reviewed in 
\S\ref{subsec:intersections}) suggests that these
quantities should have a natural geometric interpretation directly in
the context of the singular geometry; although this is not yet well
understood from a pure mathematics perspective.

We now turn to the matter content
in 4D F-theory models.  Matter fields in the low-energy theory can
arise from both localized and global features in the gauge divisor
$\Sigma$.  When a gauge divisor intersects another component of the
discriminant locus over a curve $C$, in general the fiber singularity
is enhanced over $C$, resulting in matter multiplets in the 4D theory.
In the resolved geometry these enhancements result in additional
$\P^1$ components in the fibers over $C$ (giving ``matter surfaces'').
In the M-theory picture, the matter multiplets are associated with
M2-branes wrapping these fibral curves.  When a non-abelian gauge
divisor intersects another non-abelian gauge divisor the resulting
matter is charged under both gauge groups, while intersections with
the residual discriminant locus over components not carrying a gauge
group (like the $I_1$ locus where $f, g \neq 0,\Delta = 0$) give
matter that is only charged under the single gauge factor.  There is also
``bulk'' matter in the adjoint representation supported over the full
divisor $\Sigma$.  In general, chiral matter is associated with flux
through the matter surfaces associated with the $\P^1$ fibers over matter
curves $C$.  This story is now well understood in the F-theory
literature and is reviewed in \cite{WeigandTASI}; we briefly summarize
some aspects here and return to this subject in
\S\ref{subsec:intersections} and \S\ref{subsec:matter}.
\begin{comment}
\drop{ Since
  a non-abelian gauge group appears on gauge divisor $\Sigma$, there
  is matter in the adjoint representation supported on the bulk of
  $\Sigma$. In general it includes both vector and (anti-)chiral
  multiplets. In addition, when two gauge divisors intersect with each
  other, or a gauge divisor self-intersects on a codimension-2 locus,
  the fiber singularity enhances on the locus, resulting in additional
  (anti-)chiral multiplets in other representations. Therefore, such a
  locus or curve is called matter curve $C$. It is the easiest to
  understand this in the dual M-theory picture. As the fiber becomes
  more singular on the matter curve, more blowups are needed to
  resolve the fourfold. This introduces new $\mathbb P^1$-fibers on
  the curve, and the matter states are given by M2-branes wrapping
  these fibral curves. In this paper, we only focus on the case when a
  gauge divisor self-intersects. }
\end{comment}

In many situations
the matter representations $R$ over a matter curve $C$
can be 
determined in a relatively simple way
 directly from the singular geometry \cite{KatzVafa}. First,  one
determines the vanishing orders on $C$ and associates them with a
(naive) Kodaira type, hence a larger non-abelian group $\tilde G$. 
 The adjoint representation of $\tilde G$ can then be decomposed
into
representations of the original gauge group $G$. Apart from the
adjoint of $G$ supported on the bulk of $\Sigma$, this also
includes some new representations and some singlets. These are
the matter representations supported on $C$. We denote $C_R$ as the matter
curve supporting representation $R$.
In this paper we generally avoid situations where the degrees of a codimension-2 singularity reach $(4,6)$
or higher, where the above picture breaks down, signaling the
presence of strongly coupled 
% superconformal
sectors
\cite{HeckmanMorrisonVafa,Apruzzi:2018oge}.

While determining the representations is straightforward, in 4D it is
much harder to calculate the multiplicities. In particular, they
depend on both the geometry and flux data, which are still not
fully understood. Fortunately, the calculation of chiral indices
(i.e., the difference between the numbers of chiral and anti-chiral
multiplets) has been well established (and is reviewed in
\S\ref{subsec:matter}).  The computation of the number of vector-like
chiral/anti-chiral pairs is much more subtle
\cite{Bies:2014sra,Bies:2021nje,Bies:2021xfh}.
%Such calculation is one of the main subjects
%in this paper.
When the geometric gauge
group $G$ itself is broken by flux to a smaller group $G' \subset G$,
matter can appear in various representations of $G'$ that are contained
within the representations of $G$ that may arise geometrically in the
unbroken theory.  One of the main subjects of this work is the
systematic analysis of chiral matter multiplicities for the
representations of $G'$ in such situations.  Remarkably, chiral matter
can arise for $G'$ even when there are no allowed chiral
representations of $G$ (such as for $G = E_7$).

\subsection{Tuned and rigid gauge groups}
\label{subsec:groups}

While the associated (non-abelian) gauge group factor can be easily
determined when given a singular gauge divisor, it is interesting to
consider the possible different origins of these singularities and
associated groups.  In particular, there are two main classes of gauge
group factors, namely tuned and rigid groups, which have qualitatively
different origins.

Tuned gauge groups are easily understood using the general
description of a Weierstrass model given in the previous section. 
Such gauge groups are obtained on a divisor in any base by fine-tuning (many) complex
structure moduli.
Roughly speaking,
we can do a local expansion of the
Weierstrass model around the divisor $\Sigma$:
\begin{align}
    f &= f_0+f_1 s+f_2 s^2+...\,, \nonumber \\
    g &= g_0+g_1 s+g_2 s^2+... \,,
\label{eq:heuristic-rigid}
\end{align}
where the coefficient functions live in various line bundles.  By
fine-tuning these Weierstrass coefficients $f_i,g_i$, over a divisor
whose normal bundle is not strongly negative, we can get various
orders of vanishing up to $(4,6)$.  In this way, any gauge group
factor in Table \ref{Kodaira} can be tuned over many divisors, such as
the plane $H$ in the simple base $\P^3$.

On the other
hand, many F-theory bases contain rigid gauge groups, which do not
require any fine-tuning like that described above, and are therefore
present throughout the whole set of moduli space branches associated
with elliptic fibrations over that base
\cite{MorrisonTaylorClusters,MorrisonTaylor4DClusters}.  Such rigid
gauge groups arise when a divisor $\Sigma$ has a sufficiently negative
normal bundle $N_\Sigma$; the associated strong curvature forces
sufficiently high degrees of singularity on the Weierstrass model that
a non-abelian gauge factor automatically arises on $\Sigma$.  Since
the gauge group does not depend on any moduli, there is no geometric
deformation that can break the gauge group. From the low-energy
perspective such a deformation corresponds to Higgsing, so these
groups are also called (geometrically) non-Higgsable gauge
groups. They can, however, be broken by certain types of flux
background, which we demonstrate below.  
And, when supersymmetry is broken, these groups can also be broken by
the standard Higgs mechanism by a massive charged scalar Higgs field
in the usual way.
Therefore, to avoid confusion
we refer to these gauge factors that are forced by geometry
as ``rigid'' gauge groups in this paper.  Exploration of the landscape of
allowed bases for elliptic CY threefolds and fourfolds, giving
6D and 4D F-theory models respectively, has given strong evidence that
the vast majority of F-theory bases support multiple disjoint clusters
of rigid gauge factors
\cite{TaylorWangMC,HalversonLongSungAlg,TaylorWangLandscape}.
Indeed, the only bases that do not support rigid gauge factors are
essentially the weak Fano bases, which form a tiny subset of
the full set of allowed bases (for example, for surfaces for 6D
F-theory models the only bases without rigid gauge factors are the
generalized del Pezzo surfaces, which contain no curves of
self-intersection below $-2$; among toric bases these represent only a
handful of the roughly 60,000 possible base surfaces, and for
threefold bases the weak Fano bases are an even smaller fraction of
the full set of possibilities).

The possible rigid gauge groups  in 4D F-theory models have been
completely classified \cite{MorrisonTaylor4DClusters}, in terms of
single factors and intersecting pairs of gauge factors that may arise. Unlike
% the above
gauge groups that can be realized through tuned
Weierstrass models, not all gauge groups
in Table \ref{Kodaira} can be rigid. For a single gauge factor,
the possible rigid gauge algebras are
\begin{equation}
    \mathfrak{su}(2), \mathfrak{su}(3), \mathfrak{g}_2, 
    \mathfrak{so}(7), \mathfrak{so}(8), \mathfrak{f}_4,
    \mathfrak{e}_6, \mathfrak{e}_7, \mathfrak{e}_8\,.
\end{equation}
Of these single factors, the only ones that contain $\gsm$ 
as a subgroup are
$E_8,E_7$, and $E_6$. For a product of two gauge factors, the
possible algebras are
\begin{equation}
    \mathfrak{su}(2)\oplus \mathfrak{su}(2),
\hspace*{0.1in} \mathfrak{su}(3)\oplus \mathfrak{su}(2),
\hspace*{0.1in} \mathfrak{su}(3)\oplus \mathfrak{su}(3),
\hspace*{0.1in} \mathfrak{g}_2\oplus \mathfrak{su}(2),
\hspace*{0.1in} \mathfrak{so}(7)\oplus \mathfrak{su}(2)\,.
\end{equation}
In particular, this includes the non-abelian part of $\gsm$ but,
as mentioned in Section \ref{sec:Intro}, it is hard to incorporate the remaining
$\U(1)$ in a rigid way. In this paper, we focus on the case of a
single gauge factor that contains $\gsm$. 
Formalizing the heuristic picture of (\ref{eq:heuristic-rigid}),
the
presence of a given rigid gauge factor can be easily determined by the
following analysis
\cite{MorrisonTaylor4DClusters}: we define the following divisors on
$\Sigma$ (not the base $B$)
\begin{align}
    F_k &= -4K_\Sigma+(4-k)N_\Sigma\,, \nonumber \\
    G_l &= -6K_\Sigma+(6-l)N_\Sigma\,.
\end{align}
We then determine the minimum values of $(k,l)$ such that $F_k,G_l$
are effective.  Any Weierstrass model is then forced to have vanishing
orders of at least $(k,l)$ on $\Sigma$.  When $\Sigma$ is near or
intersecting other divisors with sufficiently negative normal bundles,
this can cause a further enhancement of the gauge group factor over
$\Sigma$; for example, this effect arises in the 6D case where an
isolated curve of self-intersection $-3$ supports a rigid $SU(3)$
gauge factor, but a pair of intersecting curves with
self-intersections $(-3, -2)$ support a rigid $G_2 \times SU(2)$ group
with a minimum amount of jointly charged matter (which is insufficient
to Higgs the group down to a smaller subgroup)
\cite{MorrisonTaylorClusters}.

Rigid gauge groups are much more generic than the tuned ones in the
landscape for various reasons.  First, tuned gauge factors require
fine-tuning of moduli over any given base, while we get rigid gauge
groups automatically when the base contains divisors with reasonably
negative normal bundles.  Second, as mentioned above, most bases
contain many rigid gauge factors, so such factors are clearly
ubiquitous in the landscape.  Third, since many divisors already
support rigid gauge groups, on a generic base few (or even no)
divisors are available for tuning additional gauge factors; this
effect becomes increasingly strong as $h^{1, 1}(B)$ increases and the
number of complex structure moduli $h^{3, 1}(\hat{Y})$ (for a threefold base)
decreases. Therefore, from a statistical point of view (such as in
e.g.\ \cite{AshokDouglas,DenefDouglas}), in the absence of other
considerations not yet understood, we may expect that it is much more
likely for $\gsm$ to arise from rigid gauge groups than from simply
fine tuning over a set of divisors that do not support rigid gauge
groups, over a base such as a weak Fano threefold.

It is natural then to consider classes of models in which the SM gauge group $\gsm$ arises from a rigid gauge factor $E_6,E_7$,
or $E_8$.  While the precise abundance of these three gauge groups in
the landscape is not fully understood, it is clear that each of them
arises as a rigid gauge factor over a vast set of bases, both for 6D
and 4D F-theory models.  This abundance is most clearly understood for
6D F-theory models, where the toric bases for such models have been
completely classified \cite{MorrisonTaylorToric} and there is also
some understanding of the full set of allowed non-toric bases,
particularly at large $h^{2, 1}(\hat{Y})$
\cite{MartiniTaylorSemitoric,TaylorWangNon-toric}.  In particular,
among the 61539 toric base twofolds (including toric bases with $-9,
-10$ and $-11$ curves, which support rigid $E_8$ gauge factors and
contain (4, 6) points that must be blown up for a smooth base), 26958,
36698, 37056 bases have rigid $E_6,E_7,E_8$ factors respectively, so
more than half of all bases support each of $E_7$ and $E_8$ groups,
and 55332 ($\sim$ 90\%) contain a divisor that supports either an
$E_7$ or $E_8$ factor. 
At least for large Hodge numbers, the structure of non-toric bases is
similar, and toric bases form a good representative sample
\cite{TaylorWangNon-toric}, although it is plausible that at small Hodge numbers
non-toric bases with fewer large exceptional groups dominate.
On the other
hand, the total number of toric bases alone for 4D F-theory models is
$\mathcal O(10^{3000})$
\cite{TaylorWangMC,HalversonLongSungAlg,TaylorWangLandscape}, which is much
too large for explicit analysis. It is expected that $E_8$ is (much)
more generic than the other exceptional group factors for elliptic
CY fourfolds with toric bases, but there is no
good measure of the relative abundance between $E_7$ and $E_6$.  One
estimate of these abundances from a partial statistical analysis of
toric bases comes from a Monte Carlo analysis on blowups of $\mathbb
P^3$, without rigid $E_8$ factors or codimension-two $(4,6)$
singularities \cite{TaylorWangMC}. It is estimated that 18\% of the
bases in this study contain rigid $E_7$ factors and 26\% of them
contain rigid $E_6$ factors, but the errors in these estimates may be
large.  In general we expect that the two gauge groups have similar
relative abundance, while the overall fractions may get %slightly
smaller when $E_8$'s are included. This is the case for 6D F-theory
models: there are 24483 toric bases without rigid $E_8$'s, of which
18276 (75\%) contain rigid $E_7$ factors and 13843 (57\%) contain
rigid $E_6$ factors. 

The above estimates are focused on toric bases; since the construction
presented here gives the clearest Standard Model spectrum without
exotics for classes of non-toric bases, it is clearly desirable to
have some better estimates of how common rigid exceptional groups are
in the broader landscape of elliptic Calabi-Yau fourfolds with
non-toric bases.  For \cy fourfolds with threefold bases there are
also questions of how to statistically weight the sets of possible
fluxes and different triangulations of the base, each of which can
give exponentially large factors \cite{Wang:2020gmi} (see also
\cite{Demirtas:2020dbm} on related issues).  We leave a more detailed
analysis of these statistical questions for future work but it is
clear in any case that the rigid $E_6, E_7,$ and $E_8$ factors arise
on a vast class of F-theory bases $B$, which motivates our
consideration of SM-like constructions using these rigid gauge
factors.

\subsection{$G_4$ fluxes}
\label{subsec:flux}

Apart from the geometry of the elliptic fibration, 
further data is needed to fully
define a 4D F-theory model and determine its gauge group and
matter content. 
The structure of this extra data
is most easily understood in the
dual M-theory picture, where  the 3-form potential
$C_3$ and its field strength $G_4=dC_3$
provide extra parameters associated with a compactification. 
The degrees of freedom of $C_3$ contain continuous degrees of freedom
when $h^{2, 1} (\hat{Y})$ is nontrivial; completely incorporating the
effects of these degrees of freedom is necessary
to determine the exact matter
spectrum, which is a complex task with the current technologies, as
reviewed in \cite{WeigandTASI}. Fortunately for our
purposes, $G_4$ flux is sufficient to determine the gauge group
and chiral indices, and the tools for analyzing these aspects of the
theory are well developed.

In general, $G_4$ is a discrete flux that takes values in the fourth
cohomology $H^4(\hat Y,\mathbb R)$.
The quantization condition on $G_4$ is slightly subtle and is given by
\cite{Witten:1996md}
\begin{equation} \label{quantization}
    G_4+\frac{1}{2}c_2(\hat Y) \in H^4(\hat Y,\mathbb Z)\,,
\end{equation}
where $c_2(\hat Y)$ is the second Chern class of $\hat Y$. In
general, $c_2(\hat Y)$ can be odd (i.e., non-even), in which case the discrete
quantization of $G_4$ contains a half-integer shift.
In particular, this implies that in some cases we are
forced to turn on some flux that may cause flux breaking. This phenomenon
will be investigated 
%in \cite{oddc2}
further in a future publication.  In the
analysis here, we focus on cases where $\hat Y$ has an even $c_2$, so
this additional subtlety is irrelevant, whenever it is possible.

Next, to preserve the minimal amount of SUSY in 4D, $G_4$
must live in the middle cohomology i.e. $G_4\in H^{2,2}(\hat Y,\mathbb
R)\cap H^4(\hat Y,\mathbb Z/2)$. 
Supersymmetry also imposes the condition of primitivity \cite{Becker:1996gj,Gukov:1999ya}:
\begin{equation}
    J\wedge G_4=0\,,
\end{equation}
where $J$ is the K\"ahler form of $\hat Y$. This is
automatically satisfied when the geometric gauge group is not broken, but
not obviously satisfied when the gauge group is broken by (vertical) flux. The
interpretation of this condition is that it stabilizes some (but not all)
K\"ahler moduli; stabilizing these moduli within the K\"ahler cone
imposes additional flux constraints. This will be explained
further in Section \ref{subsec:primitive}.

We also have the D3-tadpole condition \cite{Sethi:1996es} that must be
satisfied for a consistent vacuum solution:
\begin{equation} \label{tadpole}
    \frac{\chi(\hat Y)}{24}-\frac{1}{2}\int_{\hat Y} G_4\wedge G_4=N_{D3}\in \mathbb Z_{\geq 0}\,,
\end{equation}
where $\chi(\hat Y)$ is the Euler characteristic of $\hat Y$,
and $N_{D3}$ is the number of D3-branes, or M2-branes in the
dual M-theory. To preserve SUSY and stability, we
forbid the presence of anti-D3-branes i.e. $N_{D3}\geq 0$. The
integrality of $N_{D3}$ is guaranteed by
Eq.\ (\ref{quantization}). This condition has an immediate
consequence on the sizes of fluxes. Recall the topological
formulae for CY fourfolds (see e.g. \cite{Klemm_1998}):
\begin{align}
    \chi &= 6(8+h^{1,1}-h^{2,1}+h^{3,1})\,, \nonumber \\
    h^{2,2} &= 44+4h^{1,1}-2h^{2,1}+4h^{3,1}\,,
\end{align}
where $h^{i,j}$ are the Hodge numbers. It is then clear that
$h^{2,2}>2\chi/3\gg \chi/24$. Therefore, if we randomly turn on
flux in the whole middle cohomology such that the tadpole constraint
is satisfied, a generic flux
configuration vanishes or has small magnitude in most of the
$h^{2,2}$ independent directions. As explained below, this is crucial to
figure out the preferred matter content, although we leave a more
precise and detailed analysis of these considerations to future work.

There are more flux constraints on the \emph{vertical} part
of $G_4$, such that $G_4$ dualizes to a consistent F-theory
background that preserves Poincar\'{e} invariance, which we return to below.
%This vertical flux is also our key to analyze
To analyze
flux breaking and chiral matter it is helpful
to consider
the orthogonal decomposition of the middle cohomology
\cite{Braun:2014xka}:
\begin{equation}
    H^{4}(\hat Y,\mathbb C)=H^{4}_\mathrm{hor}(\hat Y,\mathbb C)\oplus H^{2,2}_\mathrm{vert}(\hat Y,\mathbb C)\oplus H^{2,2}_\mathrm{rem}(\hat Y,\mathbb C)\,.
\end{equation}
The horizontal subspace comes from the complex structure
variation of the holomorphic 4-form $\Omega$ associated with
the CY fourfold. 
Flux in these directions has the effect of inducing a superpotential and
stabilizing complex structure moduli \cite{Gukov:1999ya}. The vertical subspace is spanned by products of harmonic
$(1,1)$-forms (which are Poincar\'e dual to divisors, denoted
by $[D_I]$)
\begin{equation}
    H^{2,2}_\mathrm{vert}(\hat Y,\mathbb C)=\mathrm{span}\left(
    H^{1,1}(\hat Y,\mathbb C)\wedge H^{1,1}(\hat Y,\mathbb C)\right)\,.
\label{eq:vertical-c}
\end{equation}
Finally, there may be components that do not belong to the
horizontal or vertical subspaces; these are referred to as
remainder flux.  While there are various types of remainder
flux, we will need the following type in
the analysis below. Consider a
curve $C_{\mathrm{rem}}\in H_{1,1}(\Sigma,\mathbb Z)$  in
$\Sigma$, such that its pushforward $\iota_*C_\mathrm{rem}\in H_{1,1}(B,\mathbb Z)$ is trivial, where
$\iota:\Sigma\rightarrow B$ is the inclusion map.
While such a
curve  cannot be realized on toric divisors on toric bases, it has
been suggested that such curves do exist on ``typical'' bases
\cite{Braun:2014xka}, so that toric geometry may be insufficiently
generic for this class of constructions; understanding this question
of typicality is an important problem for further study.
In any case, we
now restrict each $D_i$ onto $C_\mathrm{rem}$. Its Poincar\'e dual $[D_i|_{C_\mathrm{rem}}]$ is a $(2,2)$-form, but since $C_\mathrm{rem}$ cannot be obtained by intersections of base divisors, we must have
\begin{equation}
    \left[D_i|_{C_\mathrm{rem}}\right]\in H^{2,2}_\mathrm{rem}(\hat Y,\mathbb C)\,.
\end{equation}

Here we explain more about vertical flux. Combining (\ref{eq:vertical-c}) with  (\ref{quantization}) gives the
integral vertical subspace $H^{2,2}_\mathrm{vert}(\hat Y,\mathbb
R)\cap H^4(\hat Y,\mathbb Z)$. 
We focus primarily here on the vertical subspace spanned by integer
multiples of forms $[D_{I}] \wedge[D_J]$
\begin{equation}
    H^{2,2}_\mathrm{vert}(\hat Y,\mathbb Z):=\mathrm{span}_{\mathbb Z}\left(
    H^{1,1}(\hat Y,\mathbb Z)\wedge H^{1,1}(\hat Y,\mathbb
    Z)\right)\,.
\label{eq:vertical-z}
\end{equation}
While this subspace does not necessarily include all lattice points in
the full vertical cohomology $H^{2, 2}_\mathrm{vert} (\hat{Y},\C)\cap
H^4 (\hat{Y},\Z)$ of the same dimension, 
this subspace provides access to much of the interesting
physics, including the production of chiral matter and the flux
breaking mechanism we study in this paper.
The full intersection pairing on $H^4 (\hat{Y},\Z)$ is unimodular, and
in many cases there are elements of this lattice that have  components in the full vertical subspace
(\ref{eq:vertical-c}) that do not lie in (\ref{eq:vertical-z}).
Some
of these quantization issues have recently been discussed in, e.g.,
\cite{CveticEtAlQuadrillion, Jefferson:2021bid}, but various questions remain
outstanding regarding the full characterization of this quantization,
which is complicated further in connection with the possibility of non-even values of
$c_2 (\hat{Y})$.  We leave further analysis of these issues aside and
focus here primarily on the space (\ref{eq:vertical-z}) and, when
possible, on even $c_2 (\hat{Y})$. This will suffice for the examples
that we explore explicitly here.

Now we set up some notations for   vertical fluxes. We expand
\begin{equation}
    G_4^\mathrm{vert}=\phi_{IJ} [D_I]\wedge [D_J]\,,
\label{eq:g-phi}
\end{equation}
and work with integer (or possibly half-integer if $c_2$ is
odd) flux parameters $\phi_{IJ}$. Note that the
expansion depends on the choice of basis of base divisors,
which we will specify depending on context. We denote the
integrated flux as \cite{Grimm:2011fx}
%\begin{equation}
%    \Theta_{\Lambda\Gamma}=\int_{\hat Y} G_4\wedge [D_\Lambda]\wedge
%          [D_\Gamma]\,,
%\label{eq:integrated-flux}
%\end{equation}
%where $D_\Lambda,D_\Gamma$ are linear combinations of $D_I$.
%Clearly $\Theta_{\Lambda\Gamma}$ is also spanned by
 %$\Theta_{IJ}$.
\begin{equation}
    \Theta_{\Lambda\Gamma}=\int_{\hat Y} G_4\wedge [\Lambda]\wedge
          [\Gamma]\,,
\label{eq:integrated-flux}
\end{equation}
where $\Lambda,\Gamma$ are arbitrary linear combinations of $D_I$; subscripts $0, i, \alpha, \ldots$ refer to the basis divisors $D_0, D_i, D_\alpha, \ldots$.
Using the intersection
numbers on $\hat Y$, studying these
objects is turned into simple linear algebra problems. This
will be reviewed in more detail in the next subsection.

Now we are ready to write down the remaining flux constraints.
To preserve Poincar\'e symmetry after dualizing, we require
\cite{Dasgupta:1999ss}
\begin{equation} \label{PoincareSym}
    \Theta_{0\alpha}=\Theta_{\alpha\beta}=0\,.
\end{equation}
(Recall that Greek  indices $\alpha, \ldots$ correspond to divisors that are
pullbacks from the base, while Roman indices $i$ correspond to Cartan
divisors, and the index 0 refers to the global zero section of the elliptic
fibration.)
Next, if the whole geometric gauge symmetry is preserved, a necessary condition is that
\begin{equation} \label{gaugeSym}
    \Theta_{i\alpha}=0\,,
\end{equation}
for all $i,\alpha$, 
otherwise flux breaking occurs. This condition is not
sufficient when there is nontrivial remainder flux, which
will be discussed more in \S\ref{subsec:breaking}.
This is the starting point of our main results.
Note that, as we discuss further below, the condition
(\ref{PoincareSym}) for Poincar\'{e} symmetry
is unchanged when  flux breaking occurs, while (\ref{gaugeSym}) is violated.

\subsection{Intersection theory on fourfolds}
\label{subsec:intersections}

In \cite{Jefferson:2021bid}, a unified approach was developed for
organizing the relevant components of the intersection numbers on
$\hat{Y}$ into a resolution-independent structure
that conceptually simplifies the analysis of symmetry
constraints, flux breaking, and chiral matter.
The basic idea is that the intersection numbers 
\begin{equation}
M_{IJKL} = \int_{\hat{Y}} [D_I] \wedge [D_J] \wedge [D_K] \wedge [D_L] 
\label{eq:}
\end{equation}
can be organized into a matrix
\begin{equation}
M_{(IJ)(KL)} = M_{IJKL} = S_{IJ} \cdot S_{KL} \,,
\label{eq:}
\end{equation}
where the formal surface $S_{IJ} = D_I\cap D_J$ is equivalent to an element of vertical homology
$H_{2, 2} (\hat{Y},\Z)$, and ``dots'' denote the intersection
product.
In terms of this matrix, the equations
(\ref{eq:integrated-flux}-\ref{gaugeSym}), as well as the expressions
for chiral matter multiplicities in terms of $G_4$ can be expressed
simply in terms of linear algebra.

A key aspect of this perspective is that the basis  of formal surfaces
$S_{IJ}$ is redundant \cite{LinWeigandG4, Bies_2017}.  There are various equivalences between these
surfaces in homology; for example the set of such formal surfaces
$S_{\alpha \beta}$ associated with pullbacks of intersections of
divisors on the base naively has $h_{1, 1} (B) (h_{1, 1} (B) + 1)/2$
elements, whereas by Poincar\'{e} duality the number of homologically
independent curves on the base is $h_{2, 2} (B) = h_{1, 1} (B)$, so
there are at least  $h_{1, 1} (B) (h_{1, 1} (B) -1)/2$ redundant
formal surfaces $S_{\alpha \beta}$.  Such homological equivalences
between the $S_{IJ}$ correspond to null vectors of the matrix $M$.
Removing all such homological equivalences $\sim$ gives a reduced
matrix $M_{\rm red}$, which encodes the intersection product on middle
vertical homology/cohomology.  One of the key observations of
\cite{Jefferson:2021bid} is that this intersection matrix seems to
always be resolution invariant even though the quadruple intersection
numbers $M_{IJKL}$ are resolution dependent.

For an elliptic CY fourfold with a single non-abelian gauge factor, in
many cases\footnote{For most gauge groups this is the form of $M_{\rm
    red}$ when the gauge group is associated with an isolated
  singularity over a divisor in the base and there are no further
  enhanced singularities on curves intersecting that divisor.  The situation
  becomes more complicated when there are, e.g., further gauge factors
  on intersecting divisors, although codimension two enhanced
  singularities on curves can also arise in the absence of further
  gauge factors.  In a few other situations where the geometry has
  codimension three (4, 6) loci, including cases with the gauge group
  $E_7$, as well as other groups such as $SU(7)$ and $SO(12)$, there
  are extra homologically independent surfaces $S_{i j}$ that support
  flux but not conventional chiral matter fields; we avoid such
  situations here.} where there is no chiral matter the matrix $\mred$
takes the simple form
\begin{equation}
        M_\text{red} =  \begin{pmatrix}
               D_{\alpha'} \cdot K \cdot D_\alpha  & D_{\alpha'}  \cdot D_{\alpha} \cdot D_\beta  & 0 \\
              D_{\alpha' } \cdot D_{\beta'} \cdot D_{\alpha }  &0 &0 \\
            0 & 0& -\kappa^{ij}\Sigma \cdot D_\alpha \cdot D_{\alpha'} 
        \end{pmatrix} \,
\label{eq:simple-mr}
\end{equation}
 in a basis of independent vertical homology classes
$S_{0 \alpha},
S_{\alpha \beta}, S_{i \alpha}$.
Here $\kappa^{ij}$ is the inverse Killing metric of the
gauge algebra, which is also the Cartan matrix $C^{ij}$ for ADE
groups.
Note that the products here are taken in the base;
for convenience, in general we will not mention explicitly the space where
the products are taken, as the space ($\hat{Y}$ or $B$)
is already clear from context.
This is clearly resolution independent; indeed, the quadruple intersection
numbers involved in this matrix have long been  known in these cases
for general bases and gauge factors (see, e.g., \cite{Grimm:2011sk}).

Perhaps more remarkably, this resolution invariance of $M_{\rm red}$
up to a choice of basis
appears to hold even when there are homologically
nontrivial surfaces $S_{ij}$, which usually
(with the exceptions of the cases mentioned in the preceding footnote)
correspond to ``matter surfaces'' that
can support chiral matter. In this case, the general form of $M_{\rm
  red}$ in a basis of independent classes
$S_{0 \alpha},
S_{\alpha \beta}, S_{i \alpha}, S_{ij}$ is
\begin{equation}
        M_\text{red} =  \begin{pmatrix}
               D_{\alpha'} \cdot K \cdot D_\alpha  & D_{\alpha'}  \cdot D_{\alpha} \cdot D_\beta  & 0 & 0 \\
              D_{\alpha' } \cdot D_{\beta'} \cdot D_{\alpha }  &0 &0 & * \\
            0 & 0& -\kappa^{ij}\Sigma \cdot D_\alpha \cdot D_{\alpha'}& * \\
        0 & * &  * & *
        \end{pmatrix} \,.
\label{eq:mr-matter}
\end{equation}
While naively the elements marked with $*$ are resolution-dependent,
it was observed in \cite{Jefferson:2021bid} that up to an integer change
of basis, the matrices $M_{\rm red}$ given by (\ref{eq:mr-matter}) for
 distinct resolutions $\hat{Y}, \hat{Y}'$ of a singular
geometry $Y$ are equivalent in many classes of examples, and it was
conjectured that this resolution-independence always holds.
Furthermore,
given the form (\ref{eq:mr-matter}) there is a rational change
of basis under which $M_{\rm red}$ can be put in a canonical form
\begin{equation}
U^{\rm t}        M_\text{red} U =  \begin{pmatrix}
               D_{\alpha'} \cdot K \cdot D_\alpha  & D_{\alpha'}  \cdot D_{\alpha} \cdot D_\beta  & 0 & 0 \\
              D_{\alpha' } \cdot D_{\beta'} \cdot D_{\alpha }  &0 &0 & 0 \\
            0 & 0& -\kappa^{ij}\Sigma \cdot D_\alpha \cdot D_{\alpha'}& 0 \\
        0 & 0 &  0 &  
 \frac{M_{\rm phys}}{(\det \kappa)^2} 
        \end{pmatrix} \,.
\label{eq:canonical-mr}
\end{equation}
Here, $M_{\rm phys}$ is a matrix that in general
encodes the relations between fluxes and chiral matter; for any
particular choice of gauge group $G$, $M_{\rm phys}$ can be expressed
in terms of characteristic data of the gauge divisor and canonical
class of the base.  Note that because the transformation matrix
$U$ is generally rational, the appropriate lattice on which this
canonical form acts for physical flux configurations
is generally a finite index sublattice of $\Z^{N}$, where $N = h^{2,
  2}_{\rm vert} (\hat{Y})$.

One application of this framework is that we can very easily analyze
the constraint equations (\ref{PoincareSym}), (\ref{gaugeSym}) and
chiral matter from $G$-flux in a simple linear algebraic framework
using $\mred$.  In this matrix notation we can write
(\ref{eq:integrated-flux}) in the form
\begin{equation}
\Theta_{IJ} = M_{(IJ)(KL)} \phi_{KL} \,,
\label{eq:theta-matrix}
\end{equation}
where $\phi_{KL}$ is a vector of integers parameterizing the
$G$-flux as in (\ref{eq:g-phi}).  Restricting to an independent basis of (co-)homology
cycles, for example  (\ref{PoincareSym}) and (\ref{gaugeSym}) become
simple linear constraints on the flux $\phi$. 
In particular, because of the block-diagonal form of the matrix
$\mred$, (\ref{PoincareSym}) simply imposes the condition $\phi_{0
  \alpha} = \phi_{\alpha \beta} = 0$, and (\ref{gaugeSym}) imposes the
condition that $\phi_{i \alpha} = 0$ for all $i, \alpha$ 
when $M_{\rm red}$ is given by (\ref{eq:simple-mr}) and/or there are
no nonzero flux parameters $\phi_{ij}$.
 We will use this same
framework here to give a simple and unified analysis of flux breaking on
general bases.  In fact, in the process we find an elegant
correspondence between the structure of fluxes in the presence of flux
breaking and the canonical form of $\mred$ given in
(\ref{eq:canonical-mr})
for the geometric group $E_6$; we expect a similar correspondence to
hold for other groups with nontrivial matter surfaces.
Note that while the conjectured resolution-independence of
(\ref{eq:mr-matter}) has not been generally proven, we do not rely in
any significant way here on the validity of this conjecture; the
flux-breaking analysis for the group $E_7$ relies only on the form
(\ref{eq:simple-mr}), which is manifestly resolution-independent, and
for the $E_6$ analysis we use a specific resolution and associated
forms (\ref{eq:mr-matter}) and (\ref{eq:canonical-mr}).

\section{Formalism of flux breaking}
\label{sec:Formalism}

With the above tools, we can now present a general formalism for
describing flux breaking in F-theory.  While the basic ideas
underlying this process have been understood previously in the
literature \cite{WeigandTASI}, 
explicit examples of this phenomenon in F-theory
have to date not been studied in detail.  We describe here flux
breaking in a general way that makes possible a simple analysis of a
wide range of flux breaking scenarios over general bases.  We discuss
various technical points that are worth extra attention, and give some
simple examples. We will then apply the results of this section in
\S\ref{sec:SM} to build our SM-like models.

\subsection{Flux breaking}
\label{subsec:breaking}

The mechanism of gauge breaking using fluxes is
certainly not a new idea. In general, both vertical and
remainder fluxes are involved in flux breaking, giving
qualitatively different breaking patterns. The vertical flux,
at the same time, can also induce chiral matter.

Let us first study vertical flux.
Consider a non-abelian group $G$ with
its Cartan directions labelled by $i=1,2,...,\mathrm{rank}(G)$,
corresponding to the exceptional divisors $D_i$. It is well
known \cite{WeigandTASI} that if we turn on nonzero flux
\begin{equation}
    G_4^\mathrm{vert}=\sum_i \phi_{i\alpha} [D_i]\wedge[D_\alpha]\,,
\end{equation}
for a single $\alpha$ (in an arbitrary basis), $G$ is broken
into the commutant of $T=\phi_{i\alpha} T_i$ within $G$, where
the Cartan generators $T_i$ are associated with the simple
roots $\alpha_i$ i.e. in the co-root basis. The
commutant can be factorized into
$G'=H\times \U(1)^{\mathrm{rank}(G)-\mathrm{rank}(H)}$, where
$H$ does not contain any $\U(1)$ factors. The
remaining $\U(1)$'s, however, are also generically broken since the flux induces
masses to the corresponding gauge bosons through the St\"uckelberg mechanism
\cite{Grimm:2010ks,Grimm:2011tb} (see Appendix \ref{sec:Stuckelberg}).  Below we rephrase this procedure in
a more efficient language.

Recall that preserving the whole geometric gauge symmetry
requires $\Theta_{i\alpha}=0$ for
all $i,\alpha$. Now we violate some of these conditions by
turning on
some nonzero parameters $\phi_{i\alpha}$. Consider a generator $e_\beta$
corresponding to the root $\beta=-b_i\alpha_i$. We then
compute\footnote{Indices appearing twice are summed over; other
  summations are indicated explicitly.}
\begin{equation}
    [T,e_\beta]=-\phi_{i\alpha} C^{ij} b_j e_\beta\propto-\sum_j b_j\left<\alpha_j,\alpha_j\right>\kappa^{ij}\phi_{i\alpha} e_\beta\,,
\end{equation}
where $\left<.,.\right>$ is the inner product of root
vectors. The commutator vanishes, hence the generator is
preserved, only when
\begin{equation} \label{rootCondition}
    \sum_i b_i\left<\alpha_i,\alpha_i\right>\Theta_{i\alpha}=0\,,
\end{equation}
for all
$\alpha$. By Appendix
\ref{sec:Stuckelberg}, the corresponding linear combination of
Cartan generators
\begin{equation}
    \sum_i b_i\left<\alpha_i,\alpha_i\right> T_i\,,
\end{equation}
is also preserved. These
generators form the non-abelian group $H$ after breaking.
Below we will focus on ADE groups, so
$\left<\alpha_i,\alpha_i\right>$ are the same for all $i$ and
Eq.\ (\ref{rootCondition}) simply becomes
$b_i\Theta_{i\alpha}=0$ for all $\alpha$.
% \drop{If we turn on nonzero
%$\phi_{i\alpha}$ for multiple
%distinct $\alpha$'s, for simplicity we require the ratio
%$\phi_{i\alpha}/\phi_{i\alpha'}$ for nonzero
%$\phi_{i\alpha},\phi_{i\alpha'}$ to be the same for all $i$ (for each
%pair $\alpha, \alpha'$),
%such that they all give the same gauge group $H$. As we will see,
%this condition also follows from primitivity.}

The simplest example of vertical flux breaking is that we turn on $\Theta_{i'\alpha}\neq 0$
for some set of Dynkin indices $i'\in I'$ and some $\alpha$, in a
generic way such that Eq.\ (\ref{rootCondition}) is satisfied only when $b_{i'}=0$ for
all $i' \in I'$. Then
$H$ is given by removing the corresponding
nodes in the Dynkin diagram of $G$. The simple roots of $H$ are
directly descended from $G$ and are given by $\alpha_{i\notin I'}$.  We will focus on this kind of breaking below.
\begin{comment}
If $\Theta_{i'\alpha}\neq 0$ for some
$i',\alpha$, the corresponding Cartan gauge bosons will get
St\"uckelberg masses (see Appendix \ref{sec:Stuckelberg}) and the gauge group $G$ breaks to $H$. The
non-abelian part of $H$ is given by removing the corresponding
nodes in the Dynkin diagram of $G$. The simple roots of $H$ are
directly descended from $G$ and are given by $\alpha_{i\notin i'}$. From now on we denote the broken directions as $i'$.
The correspondence between the two descriptions is as
follows. Consider the commutator
\begin{equation}
    [T,e_j]=\sum_i \phi_{i\alpha} C_{ij} e_j
    \propto\sum_i \kappa^{-1}_{ij}\phi_{i\alpha}e_j\,,
\end{equation}
which vanishes if $\Theta_{j\alpha}=0$ for all $\alpha$,
otherwise equals to $\Theta_{j\alpha'}e_j$ for some $\alpha'$.
Here $e_j$ is the generator corresponding to the simple root
$\alpha_j$. Therefore, the commutant of
$T$ in $G$ has roots spanned by $\alpha_j$ with
$\Theta_{j\alpha}=0$ for all $\alpha$.
\end{comment}

The statements for remainder flux are similar. If we turn on
\begin{equation}
    G_4^\mathrm{rem}=[c_i D_i|_{C_\mathrm{rem}}]\,,
\end{equation}
for some $C_\mathrm{rem}$ satisfying the property mentioned
in \S\ref{subsec:flux}, $G$ is broken into the commutant of
$T=c_i T_i$ within $G$. The difference is that the remainder
flux does not turn on any $\Theta_{i\alpha}$, so there is no
St\"uckelberg mechanism and all the $\U(1)$ factors in 
the commutant are preserved. In other words, breaking using
remainder flux never decreases the rank of the gauge group,
while breaking using vertical flux always decreases the rank.
In general, when both types of fluxes are turned on, only the
intersection of the two commutants are preserved. As a
result, a wide variety of breaking patterns can be
constructed using combinations of these fluxes. Note that
when $G$ is a rigid gauge group, $\Sigma$ is a rigid divisor
and supports remainder flux breaking only when embedded into
a non-toric base.
This follows because for a toric base $B$, toric divisors span the
cone of effective divisors, so any rigid effective divisor $\Sigma$ is
toric, and toric curves in a toric $\Sigma$ span $h^{1,1} (\Sigma)$.

So far we have focused on the non-abelian part of the broken
gauge group, while there can also be $\U(1)$ factors remaining. There
are two ways to get $\U(1)$'s in our formalism. The first way
is, obviously, breaking $G$ with remainder flux in which all
the $\U(1)$'s in the commutant are preserved. It is also
possible to get $\U(1)$'s with vertical flux. By imposing
$p_i\Theta_{i\alpha}=0$ for all $\alpha$ and some $p_i$, the
linear combination of Cartan generators $T_p=p_i T_i$ is
preserved. By choosing $p_i$ such that $p_i \alpha_i$ (modulo the preserved roots) is not
along a root of $G$, there is no additional root to be preserved, so $T_p$
corresponds to an extra $\U(1)$ factor instead of a part of
$H$. Note that the $\U(1)$'s induced by vertical flux are
always ``exotic'': a $\U(1)$ that coincides with a root of $G$
must be obtained through remainder instead of vertical flux,
otherwise the $\U(1)$ enhances to a part of $H$.
\begin{comment}
On the
other hand, if there is a linear relation between
$\Theta_{i'\alpha}$'s with different $i'$ i.e.
$c_{i'}\Theta_{i'\alpha}=0$ for all $\alpha$, the corresponding
Cartan direction $c_{i'}T_{i'}$ becomes massless again and is a
part of $H$. In practice, we choose the non-abelian part of $H$
by choosing the Dynkin nodes to be removed, and the abelian
part of $H$ by imposing these linear relations.
\end{comment}

There is an additional 
subtlety from vertical flux breaking. Let the
$\alpha$'s giving homologically independent $S_{i\alpha}$ be
$\alpha_1,\alpha_2, \ldots,\alpha_r$.
\begin{comment}
By basic linear algebra,
no matter how we turn on $\phi_{i\alpha}$, there must be at
least $\left(\mathrm{rank}(G)-r\right)$ linear relations in the form of
Eq.\ (\ref{rootCondition}). Therefore, we must have
\end{comment}
From the above breaking rules, we see that the
difference $\mathrm{rank}(G)-\mathrm{rank}(G')$ is
given by the rank of the  ($r$  $\times$
$\mathrm{rank}(G)$)
matrix $\Theta_{(\alpha_a)(i)}$ (where $a$ and $i$
are the indices for rows and columns respectively).
As we will show in Section \ref{subsec:primitive},
to satisfy primitivity the rank of the matrix is constrained to be
 at most $r-1$.
Therefore, we get a lower bound on $r$ for given $G$ and $G'$:
\begin{equation} \label{rlowerbound}
    r\geq\mathrm{rank}(G)-\mathrm{rank}(G')+1\,.
\end{equation}
In particular, we must have a sufficiently large number $r$
of $\alpha$'s giving independent cycles $S_{i \alpha}$
(associated with independent curves in $\Sigma$ that are also
independent in $B$) in order
to get a desired $G'$. This condition imposes constraints on
the possible geometries that support a given vertical flux breaking.
\begin{comment}
Apart from the linear relations
we impose, we may be able to construct some \emph{accidental}
ones that exist but are not imposed intentionally. In other
words, the St\"uckelberg mass matrix (see Appendix
\ref{sec:Stuckelberg}) has rank at most $h^{1,1}(\Sigma)$,
which may be smaller than our desired number of massive
directions. This leads to extra $\U(1)$'s and makes $H$
deviate from our desired one. To avoid so, we must require
\begin{equation}
    \mathrm{rank}(G)-\mathrm{rank}(H)\geq h^{1,1}(\Sigma)\,.
\end{equation}
\end{comment}

\subsection{Chiral matter and matter surfaces}
\label{subsec:matter}

Apart from breaking the gauge group, the \emph{vertical} flux
can also induce chiral matter. The famous
index formula states that for a weight $\beta$ in
representation $R$, its chiral index $\chi_\beta$ is \cite{Braun_2012,Marsano_2011,KRAUSE20121,Grimm:2011fx}
\begin{equation} \label{ordinarychi}
    \chi_\beta=\int_{S(\beta)}G_4^\mathrm{vert}\,,
\end{equation}
where $S(\beta)$ is called the matter surface of $\beta$. When
$R$ is localized on a matter curve $C_R$, $S(\beta)$ is the
fibration of the blowup $\mathbb P^1$ corresponding to $\beta$
over $C_R$. When $G$ is not broken, the vanishing of all
$\Theta_{i\alpha}$ guarantees that all $\beta$ in $R$ give the
same $\chi_R$. When $G$ is broken to $G'$, $R$ decomposes into
different irreducible representations $R'$ in $G'$ and the above is no
longer true. Instead, we need that all $\beta'$ in $R'$ give
the same $\chi_{R'}$.

This can be seen as follows. Since weights differ by roots,
given a weight $\beta$ in $R$ of $G$, it is useful to
expand
$\beta=-b_i\alpha_i$. Hence we can decompose its matter surface
$S(\beta)$ as \cite{WeigandTASI}
\begin{equation} \label{Sdecompose}
    S(\beta)=S_0(R)+b_i\left.D_i\right|_{C_R}\,,
\end{equation}
where $S_0$ only depends on $R$ but not $\beta$. We will prove
this decomposition below. When $G$ is not broken, $\chi_R$ is
calculated using $S_0$. The Poincare dual $[S_0(R)]$ is the corresponding
%matter surface
flux that gives chiral matter without breaking
$G$ when $G$ supports chiral matter.  As we will see more explicitly below, $S_0 (R)$ and its Poincar\'{e} dual correspond to the last row/column of (\ref{eq:canonical-mr}). We will now focus on the
second term of (\ref{Sdecompose}) and determine $S_0$ later. Matter curves in general
can be written as
\begin{equation}
    C_R=\Sigma\cdot(p_R K_B+q_R \Sigma)\,,
\end{equation}
where $p_R,q_R$ are some (integer) coefficients. Then,
\begin{equation} \label{partialchiR'}
    \int_{S(\beta)}G_4^\mathrm{vert}=\int_{S_0(R)}G_4^\mathrm{vert}+b_i\int_{\hat Y} G_4^\mathrm{vert}\wedge[D_i]\wedge\pi^*[p_R K_B+q_R \Sigma]\,.
\end{equation}
The second term is a linear combination of $\Theta_{i\alpha}$
and we can replace the $i$ summation with $i' \in I'$ since the other terms vanish. Since the
weights of $R'$ differ by combinations of $\alpha_{i\notin I'}$
only, each set of $b_{i'}$ gives a representation $R'$, and Eq.\ (\ref{partialchiR'}) is the same for all weights of $R'$. In
general, different $b_{i'}$ and different $R$ can give  rise to the same irreducible representation
$R'$. We must sum over these contributions to get the complete $\chi_{R'}$.
Applying Eq.\ (\ref{ordinarychi}), we get
\begin{equation} \label{fullchiR'}
    \chi_{R'}=\sum_R \sum_{b_{i'}} \left(\int_{S_0(R)}G_4^\mathrm{vert}+b_{i'}\left(p_R \Theta_{i'K_B}+q_R \Theta_{i'\Sigma}\right)\right)\,.
\end{equation}
This is our main tool to calculate chiral indices in models
with flux breaking. An important feature that can be seen here
is that $\chi_{R'}$ for complex $R'$ can be nontrivial even if
$R$ is non-complex. In other words, there can be chiral matter
after flux breaking even if $G$ does not support chiral matter.
This formula passes several consistency checks, such as $\chi_{\bar R'}=-\chi_{R'}$, since taking the conjugate representation flips
all contributions in Eq.\ (\ref{fullchiR'}) to opposite signs.
Moreover, in all examples we will see, anomaly cancellation is
preserved after the breaking as long as the flux constraints
are satisfied.

So far, we have been focusing on matter localized on curves. On
the other hand, adjoint matter lives on the bulk of $\Sigma$
and matter curves or surfaces for this representation are not well-defined.
Nevertheless, it has been shown that adjoint matter can also
become chiral after flux breaking, and the chiral indices are
given by setting $S_0(\mathrm{Adj})=0$ and replacing $C_R$ by
$K_\Sigma$ \cite{Bies:2017fam}. By the adjunction formula,
$K_\Sigma=\Sigma\cdot(K_B+\Sigma)$ and we should set
$p_\mathrm{Adj}=q_\mathrm{Adj}=1$.

It may sound strange that $K_{\Sigma}$ directly appears in
$\chi$, while in 6D F-theory models it is well-known that the
number of adjoint hypermultiplets is the genus
$g=\left(K_{\Sigma}+2\right)/2$ \cite{Witten:1996qb}. Should there also be such
a shift in the 4D formula? In fact, we should compare the formula
with the Dirac index of adjoint matter in 6D instead. In 6D
$\mathcal{N}=1$ SUSY, each vector multiplet contains two
$(0,1/2)$ spinors, while each hypermultiplet
(two half-hypermultiplets) contains two $(1/2,0)$ spinors.
Since there is one vector multiplet and there are $g$ hypermultiplets,
the Dirac index in 6D is indeed $2g-2=K_{\Sigma}$.

Now we return to the determination of $S_0$. Since the
nontrivial $\Theta$ are $\Theta_{ij}$ and $\Theta_{i\alpha}$,
we only need the $S_{ij}$ and $S_{i\alpha}$ components in
$S_0$. The $S_{ij}$ components, if they exist, give chiral matter
even when $G$ is not broken.  A useful indirect procedure to determine such
components has been established,  through the
matching of Chern-Simons (CS) terms in M/F-theory duality \cite{Grimm:2011sk,Grimm:2011fx,Cvetic:2012xn}. To
be precise, in the 3D M-theory dual, $\Theta_{ij}$ are the
classical CS couplings appearing in the 3D effective action. These
match with the one-loop corrected CS couplings in the 4D
F-theory when compactified (additionally) on a circle. The
charged fermions running through the loop relate the couplings
to chiral indices. As a result, we can establish relations in
the form of $\chi_R=x_R^{ij}\Theta_{ij}$, where $x_R^{ij}$ are
some coefficients.
We refer to \cite{Jefferson:2021bid} for more details. Note that
this determines the $S_{ij}$ components in $S_0$, which are
sufficient when $G$ is not broken. To include the $S_{i\alpha}$
components for the broken case, we make the following ansatz:
\begin{equation}
    S_0(R)=x_R^{ij}D_i\cdot D_j+D_i\cdot D_R^i\,,
\end{equation}
where $D_R^i$ is some linear combination of $D_\alpha$. Now we
determine $D^i_R$. First, we choose a base divisor $D'$ such
that it intersects $C_R$ only once i.e. $C_R\cdot D'=1$. Then
by definition, the fibral curve $C_\beta$ corresponding to
weight $\beta$ is
\begin{equation}
    C_\beta=S(\beta)\cdot D'=S_0(R)\cdot D'+b_i\mathbb P^1_i\,,
\end{equation}
where $\mathbb P^1_i$ is the fibral curve in $D_i$. Now we must have
\begin{equation}
    D_i\cdot C_\beta=\beta_i\,,
\end{equation}
where $\beta_i=-C_{ij} b_j$ are the components of $\beta$ in a basis of
fundamental weights.
By $D_i\cdot \mathbb P^1_j=-C_{ij}$, we see that the
second term in Eq.\ (\ref{Sdecompose}) gives all the weights,
and the condition reduces to simply
\begin{equation}
    S_0(R)\cdot D_i\cdot D'=0\,.
\end{equation}
All intersection numbers in the above involve triple
intersections of $\Sigma,D'$, and some other classes on the
base. Since we have the freedom to choose $D'$ as long as it is
properly normalized, the $\mathrm{rank}(G)$ constraints
determine $\Sigma\cdot D^i_R$ in terms of other known
classes, namely $\Sigma^2,\Sigma\cdot K_B$. This is equivalent to
determining $D_i\cdot D^i_R$ since only $\Sigma\cdot D^i_R$
appears in its intersection numbers. Therefore, $S_0(R)$ has
been fixed. Notice that these constraints also mean that $S_0(R)$
must live in the directions of $M_{\mathrm{phys}}$, confirming that this surface and the associated Poincar\'{e} dual flux correspond to the final row/column of (\ref{eq:canonical-mr}) as asserted above; in fact this conclusion can also be arrived at directly from the observation that the Poincar\'{e} dual $[S_0 (R)]$ is the only flux direction that preserves Poincar\'{e} and gauge symmetry. The
block-diagonal form of $M_{\mathrm{red}}$ then implies that we
can always separate the chiral indices into contributions
preserving $G$, and those induced by flux breaking. These
correspond to the two terms in Eq.\ (\ref{Sdecompose}), hence
give a resolution-independent description of matter surfaces.
This also recovers the statement that chiral matter in $G$ is
induced by flux along the Poincar\'e dual $[S_0(R)]$ \cite{Marsano_2011}.
We will explicitly demonstrate these  relations in Section \ref{subsec:e6}.

It is useful to have a simple result from the above procedure.
For non-complex $R$, it is clear that the procedure gives
trivial $S_0(R)$. In particular, for $G$ not supporting chiral
matter, $S_0(R)$ is always absent.

\subsection{Primitivity}
\label{subsec:primitive}

The gauge-breaking flux must also satisfy various flux
constraints discussed in Section \ref{subsec:flux}. Interestingly, the \emph{vertical} flux
we turn on does not automatically
satisfy primitivity ($J\wedge G_4=0$). Extra attention must be paid and we will see
that primitivity leads to additional flux constraints.

It is useful to first review the basics of the K\"ahler form $J$.
The volume of a complex $d$-dimensional submanifold $\mathcal M^d$ in $\hat Y$ is given by
\begin{equation}
    \mathrm{vol}(\mathcal M^d)=\int_{\mathcal M^d} J^d\,.
\end{equation}
The K\"ahler cone is then the cone of $J$ giving positive volumes. We can expand $J$ of $\hat Y$ in the K\"ahler cone as
\begin{equation}
    J=t^0[D_0]+t^\alpha[D_\alpha]+t^i[D_i]\,,
\end{equation}
where the K\"ahler moduli $t$ are
restricted to the positive K\"ahler cone.
%positive and we have chosen the
%K\"ahler basis on the base for $D_\alpha$.
So far we have focused
on the resolved manifold $\hat Y$, which is on the M-theory
Coulomb branch where $G$ is broken into
$\U(1)^{\mathrm{rank}(G)}$. To take the F-theory limit and
restore the whole $G$, we need to shrink the fibers to zero
volume, while keeping the (pullbacks of the) base divisors at finite
volumes. Note that $t^0$ and $t^i$ measure the elliptic and
exceptional fiber volumes respectively. Therefore, we need to
send $t^0$ and $t^i$ to zero and scale up $t^\alpha$. To be
precise, the limit can be done by the following rescaling \cite{Grimm:2010ks,Bonetti:2011mw}:
\begin{equation}
    t^0\rightarrow \epsilon t^0\,,\quad t^\alpha\rightarrow \epsilon^{-1/2} t^\alpha\,,\quad t^i\rightarrow \epsilon^{3/2} t^i\,,
\end{equation}
where the limit is now $\epsilon\rightarrow 0$. Therefore, we
only need to consider $J\rightarrow\pi^* J_B=t^\alpha[D_\alpha]$ when studying primitivity.

First we recall how primitivity is satisfied when $G$ is not
broken. By Eq.\ (\ref{PoincareSym}) and (\ref{gaugeSym}), we
have $\Theta_{I\alpha}=0$ for all $\alpha$. This already
guarantees $J\wedge G_4=0$ and primitivity is automatically
satisfied. It is also clear that nonzero $\Theta_{i\alpha}$
breaks the above argument and primitivity is not always
satisfied. In particular, generically we have
\begin{equation}
    \int_{\hat Y} [D_i]\wedge J\wedge G_4=t^\alpha \Theta_{i\alpha}\neq 0\,.
\end{equation}
The above vanishes only for specific values of $t^\alpha$. The
interpretation is that by turning on gauge-breaking flux, some
K\"ahler moduli are stabilized (but not all, as an overall
rescaling of $t^\alpha$ also satisfies the constraint).

On the other hand, not all choices of nonzero
$\Theta_{i\alpha}$ can stabilize the K\"ahler moduli within the
K\"ahler cone. As a first step, one necessary condition for consistent
stabilization is that the flux should give a positive tadpole
$\int_{\hat Y}G_4\wedge G_4>0$, which is already not always
true for gauge-breaking vertical flux. From the form of
$M_{\mathrm{red}}$, the sign of the tadpole is determined by the
triple intersection form $\Sigma\cdot D_\alpha\cdot D_\beta$
on the base. If this is positive semidefinite, the flux always
gives a nonpositive tadpole, hence is not ever consistent. Although
the intersection forms for most geometries of $\Sigma$ have both
positive and negative directions, $\Sigma$ cannot be as simple
as $\mathbb P^2$. When the
tadpole can be positive, during vertical flux breaking we must turn on
some gauge-breaking flux along negative
directions in the intersection form of $\Sigma$.

We must stress again that having  a positive tadpole is not a
sufficient condition for primitivity. Here we show  how
primitivity leads to additional flux and geometric constraints. We consider
\begin{equation}
    t^\alpha \Theta_{i\alpha}=0\Rightarrow t^\alpha \Sigma\cdot D_\alpha \cdot D_\beta \phi_{i\beta}=0\,,
\end{equation}
for all $i$.  
%\drop{This is satisfied by the condition that
%$\phi_{i\alpha}/\phi_{i\alpha'}$ is the same for all
%$i$, but} 
Notice that the solutions of
$t^\alpha$ live in the left null space of the matrix
$\Theta_{(\alpha_a)(i)}$. Therefore to have
nontrivial solutions of $t^\alpha$, the rank of the
matrix must be less than $r$, i.e., at most $r-1$,
where we recall that $r$ is the number of  $\alpha$'s giving
homologically independent cycles $S_{i \alpha}$.
This leads to Eq.\ (\ref{rlowerbound}).  The
positivity of $t^\alpha$ is more subtle.
%For an arbitrary $i$,
%we now decompose the above into the Mori basis $l_\alpha$ on
%the base (which is dual to the K\"ahler basis $D_\alpha$).
In the simplest cases where the Mori cone (dual to the K\"ahler cone) is generated by $h^{1,1} (B)$ basis curves $l_\alpha$, we can decompose the above into this basis.
That is, for any $i$ we have $\phi_{i\beta}\Sigma\cdot D_\beta=m_{i\alpha} l_\alpha$
for some coefficients $m_{i\alpha}$. Then the primitivity constraint
is simply
\begin{equation}
    t^\alpha m_{i\alpha}=0\,.
\end{equation}
Therefore to have positive $t^\alpha$, we must have at least a
pair of $m_{i\alpha}$ with opposite signs. This imposes some sign
constraints on $\phi_{i\alpha}$ as demonstrated below in some specific
cases.

\subsection{Simple $\SU(N)$ models}
\label{subsec:SUNex}

So far we have presented the general formalism of flux
breaking. To see how it works, it is useful to  illustrate with some simple
examples involving vertical flux breaking of $G=\SU(N)$ to
$G'=\SU(N-1)$ with no extra $\U(1)$.
We focus on checking the formalism using anomaly cancellation, which is automatically
achieved in all examples below as a consequence of (\ref{fullchiR'}). Interestingly, this is a
result from nontrivial cancellation between the matter
representations in $G$. To focus on the effect of flux
breaking, we do not include any chiral matter in the unbroken
models. In other words, we focus on the second term in Eq.\ (\ref{fullchiR'}), which should satisfy anomaly cancellation
on its own as discussed.
In each case, we turn on $\Theta_{i' \alpha}$ for $i' = N -1$, to
break the Dynkin diagram $A_{N -1}$ to $A_{N -2}$.

\begin{itemize}
    \item $\SU(3)\rightarrow \SU(2)$
\end{itemize}

Since $\SU(2)$ does not support any chiral matter, all chiral
indices should vanish. Indeed, all $\SU(2)$ representations
come from pairs of opposite $b_{i'}$. For example, the $\SU(3)$
adjoint $\mathbf 8$ gives two copies of the $\SU(2)$
fundamental representation
$\mathbf 2$ with $b_2=\pm 1$. The $\SU(3)$ fundamental $\mathbf 3$ gives a $\mathbf 2$ with $b_2=1/3$ which is nonzero, but we
also have the $\SU(3)$ antifundamental $\bar{\mathbf 3}$ giving
another $\mathbf 2$ with $b_2=-1/3$. Eq.\ (\ref{fullchiR'})
then implies that $\chi_{\mathbf 2}=0$. In general, such
a cancellation holds for any non-complex $R'$. Note that a
single $R$ may have nonzero contribution to $\chi_{R'}$,
although it must get cancelled. This shows that only the total
$\chi_{R'}$ is a physical quantity.

\begin{itemize}
    \item $\SU(4)\rightarrow \SU(3)$
\end{itemize}

$\SU(3)$ has complex representations such as $\mathbf 3$, but a
generic $\SU(3)$ F-theory model only contains $\mathbf 8,\mathbf 3,\bar{\mathbf 3}$, and $\chi_{\mathbf 3}=0$ is
required by anomaly cancellation. Interestingly, this is more
nontrivial from the $\SU(4)$ perspective. Consider a generic
$\SU(4)$ model, which contains the representations
$\mathbf{15},\mathbf{6},\mathbf{4}$ and the conjugates. For
the latter two, the matter curves are
$C_{\mathbf{6}}=-\Sigma\cdot K_{B}$ and
$C_{\mathbf{4}}=-\Sigma\cdot\left(8K_{B}+4\Sigma\right)$. The
three representations all break to $\mathbf{3}$ with
$b_3=1,-1/2,1/4$ respectively. Be careful here to recall that
$C_{\mathbf{6}}$ actually contains two copies of $\mathbf{6}$.
Now we have
\begin{equation}
    \sum_R b_3^R C_R=\Sigma\cdot\left(1\cdot\left(K_{B}+\Sigma\right)+2\cdot\left(-\frac{1}{2}\right)\cdot\left(-K_{B}\right)+\frac{1}{4}\cdot\left(-8K_{B}-4\Sigma\right)\right)=0\,.
\end{equation}
Eq.\ (\ref{fullchiR'}) then implies that $\chi_{\mathbf 3}=0$.
We see that anomaly cancellation has become a cancellation
%between
that involves both weights and classes of matter curves.

\begin{itemize}
    \item $\SU(5)\rightarrow \SU(4)$
\end{itemize}
The calculation is similar. A generic $\SU(5)$ model contains
$\mathbf{24},\mathbf{10},\mathbf{5}$ and the conjugates. The
matter curves are $C_{\mathbf{10}}=-\Sigma\cdot K_{B}$ and
$C_{\mathbf{5}}=-\Sigma\cdot\left(8K_{B}+5\Sigma\right)$. The
three representations all break to $\mathbf{4}$ with
$b_4=1,-3/5,1/5$ respectively. Then,
\begin{equation}
    \sum_{R}b_4^RC_{R}=\Sigma\cdot\left(1\cdot\left(K_{B}+\Sigma\right)+\left(-\frac{3}{5}\right)\cdot\left(-K_{B}\right)+\frac{1}{5}\cdot\left(-8K_{B}-5\Sigma\right)\right)=0\,.
\end{equation}
Therefore, $\chi_{\mathbf{4}}=0$ as required by anomaly cancellation.

\begin{itemize}
    \item $\SU(6)\rightarrow \SU(5)$
\end{itemize}
This is a more interesting example since $G'$ now supports
chiral matter. We show that flux breaking can induce chiral
matter satisfying anomaly cancellation, even if there is no
chiral matter in the unbroken phase. A generic $\SU(6)$
model contains $\mathbf{35},\mathbf{15},\mathbf{6}$. The
matter curves are $C_{\mathbf{15}}=-\Sigma\cdot K_{B}$ and
$C_{\mathbf{6}}=-\Sigma\cdot\left(8K_{B}+6\Sigma\right)$. All
three representations break to $\mathbf{5}$ with
$b_5=1,-2/3,1/6$ respectively, while $\mathbf{15}$ also breaks
to $\mathbf{10}$ with $b_5=1/3$. Using Eq.\ (\ref{fullchiR'}),
we get
\begin{equation}
    \chi_{\mathbf{5}}=-\chi_{\mathbf{10}}=\frac{1}{3}\Theta_{iK_{B}}\,.
\end{equation}
Therefore, the chiral indices become nontrivial, and the
anomaly cancellation condition $\chi_{\mathbf{5}}=-\chi_{\mathbf{10}}$
is satisfied. Note that despite the presence of  the factor of
$1/3$, the flux constraints must guarantee integer chiral
indices.

\section{Standard Model structure from $E_7$ flux breaking}
\label{sec:SM}

We are now ready to discuss the breaking $E_7\rightarrow\gsm$,
which leads to the SM gauge group and exact chiral spectrum from a
gauge group ubiquitous in the landscape. This is the main
result of \cite{Li:2021eyn}, but here we provide more details. In
particular, we discuss more about different embeddings of
$\gsm$ into $E_7$ and solve the flux constraints more
generally. As shown in Section
\ref{sec:othergroups}, all these results can be
easily generalized to $E_6$ flux breaking.

\subsection{Embeddings of the gauge group $\gsm$}
\label{subsec:embedding}

As a first step, here is a general picture of $E_7$ flux
breaking. A generic $E_7$ model contains the adjoint
$\mathbf{133}$ and the fundamental $\mathbf{56}$, with matter
curve $C_{\mathbf{56}}=-\Sigma\cdot(4K_B+3\Sigma)$. In
particular, there are models with only $\mathbf{133}$ but no
$\mathbf{56}$, so the two representations should satisfy
anomaly cancellation separately under flux breaking, unlike
the $\SU(N)$ models. The number of independent sets of chiral
matter induced by vertical flux breaking depends on the embedding of
$\gsm$ into $E_7$. Usually, the number allowed by anomaly
cancellation is (much) less than the number of independent flux
parameters we
turn on. Therefore, many flux parameters contribute to a
single chiral index, naturally leading to a small number of
generations as demonstrated below. More surprisingly, the
chiral matter induced may not realize all the independent sets
allowed by anomaly cancellation, unlike the situation for generic
chiral matter representations in universal tuned $\gsm$ models without
flux breaking \cite{Jefferson:2021bid}.

Below we see two examples of $\gsm$ embeddings into $E_7$. The first one
gives SM chiral matter as the only allowed matter spectrum.
The second one gives exotic chiral matter (defined below)
which is only part of the spectrum allowed by anomaly
cancellation.

\subsubsection{Standard Model chiral matter}
\label{ssubsec:SMchiral}

\begin{figure}[t]
\centering
\includegraphics[width=0.5\columnwidth]{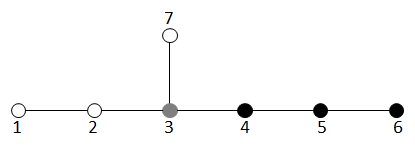}
\caption{The Dynkin diagram of $E_7$. The Dynkin node labelled
$i$ corresponds to the exceptional divisor $D_i$. The solid
nodes are the ones we break to get the Standard Model gauge
group and chiral matter. Node 3 (in gray) is broken by
remainder flux while the others are broken by vertical flux.}
\label{dynkine7}
\end{figure}

Here we consider embeddings of $\gsm$ that lead to SM chiral
matter. First, the embedding of the non-abelian part i.e.
$\SU(3)\times\SU(2)$ is unique up to $E_7$ automorphisms,
when we restrict to root embeddings,
see
Appendix \ref{sec:embeddingcount}.  Without loss of
generality, we put the non-abelian part in nodes $1,2,7$ in
the Dynkin diagram, see Figure \ref{dynkine7}. Then, there
are 4 choices of $\U(1)$ (with generator $T_Y=Y_i T_i$) in $\gsm$ that
play the role of hypercharge and give only SM chiral matter (see Appendix \ref{sec:embeddingcount}):
\begin{align} \label{hyperchargeChoice}
    Y_i=&-(1/3,2/3,1,0,0,0,1/2)\,,\quad-(1/3,2/3,1,1,0,0,1/2)\nonumber \\
    &-(1/3,2/3,1,1,1,0,1/2)\,,\quad-(1/3,2/3,1,1,1,1,1/2)\,.
\end{align}
In fact, however, these are also equivalent under
automorphisms, and
for all choices $Y_{i=3,4,5,6}$ coincides with a
root of $E_7$ that expands the group to $SU(5)$, so the hypercharge must be obtained through
remainder flux. Moreover, vertical flux is also necessary for chiral
matter. For simplicity, we will focus on the
first choice of $Y_i$, while other choices give similar
results. Therefore, the proposal is that we first break $E_7$
down to an intermediate $\SU(5)$ with vertical flux, then
obtain $\gsm$ using hypercharge flux, in parallel with
earlier work on tuned SU(5) GUT
models \cite{Donagi:2008ca,BeasleyHeckmanVafaI,BeasleyHeckmanVafaII,DonagiWijnholtGUTs,Blumenhagen:2009yv,Marsano:2009wr,Grimm:2009yu,KRAUSE20121,Braun:2013nqa}. As mentioned in
\S\ref{subsec:breaking}, the construction can be done on
typical but non-toric bases supporting rigid $E_7$ factors.

Following this approach, we first break $E_7$ down to $\SU(5)$
by turning on nonzero $\Theta_{i'\alpha}$ for $i'=4,5,6$ and
some $\alpha$, see Figure \ref{dynkine7}. Then we further
break $\SU(5)$ down to $\gsm$ by turning on the hypercharge flux:
\begin{equation} \label{hyperchargeflux}
    G_4^\mathrm{rem}=\left[D_Y|_{C_\mathrm{rem}}\right]\,,
\end{equation}
for some $C_\mathrm{rem}$, where $D_Y=2D_1+4D_2+6D_3+3D_7$ is
the exceptional divisor corresponding to the hypercharge
generator from the first choice of $Y_i$. This remainder flux
further breaks node 3 in Figure \ref{dynkine7} and gives $\gsm$.

\begin{comment}
For
simplicity, we only focus on the case where the non-abelian
part of $\gsm$ is produced by breaking the Dynkin nodes
$i'=3,4,5,6$, see Figure \ref{dynkine7}. Hence we turn on nonzero $\Theta_{i'\alpha}$ for $i'=3,4,5,6$.
\end{comment}

Since only the vertical flux induces chiral matter, we can analyze
the matter content by breaking $E_7 \rightarrow SU(5)$, where
the $\mathbf{56}$ breaks into a combination of $\mathbf{5},
\mathbf{10}$, uncharged singlets and conjugate representations, and $\mathbf{133}$
includes these as well as the adjoint $\mathbf{24}$.  Since the
adjoint is non-chiral, the only chiral representations we expect for
$\gsm$ after the final breaking by remainder flux
are the Standard Model representations 
\begin{equation}
    \left(\mathbf{3},\mathbf{2}\right)_{1/6}\,,\quad
    \left(\mathbf{3},\mathbf{1}\right)_{2/3}\,,\quad
    \left(\mathbf{3},\mathbf{1}\right)_{-1/3}\,,\quad
    \left(\mathbf{1},\mathbf{2}\right)_{1/2}\,,\quad
    \left(\mathbf{1},\mathbf{1}\right)_{1}\,.
\end{equation}

As mentioned above, we expect anomaly cancellation separately from the
matter arising from the $\mathbf{56}, \mathbf{133}$ of $E_7$.
Using Eq.\ (\ref{fullchiR'}), we
indeed get SM chiral matter from vertical flux and the $\mathbf{56}$ with
\begin{equation} \label{chi56}
    \chi_{\left(\mathbf{3},\mathbf{2}\right)_{1/6}}^{\mathbf{56}}=\frac{1}{2}\left(3\Theta^{\mathbf{56}}_4+2\Theta^{\mathbf{56}}_5+\Theta^{\mathbf{56}}_6\right)\,,
\end{equation}
where $\Theta^{\mathbf{56}}_i=-4\Theta_{iK_{B}}-3\Theta_{i\Sigma}$.
Similarly,
$\mathbf{133}$ also gives SM chiral matter with
\begin{equation} \label{chi133}
    \chi_{\left(\mathbf{3},\mathbf{2}\right)_{1/6}}^{\mathbf{133}}=-\left(3\Theta^{\mathbf{133}}_4+2\Theta^{\mathbf{133}}_5+\Theta^{\mathbf{133}}_6\right)\,,
\end{equation}
where $\Theta^{\mathbf{133}}_i=\Theta_{iK_{B}}+\Theta_{i\Sigma}$. We see that only certain linear combinations of
$\Theta_{i'\alpha}$ appear in the chiral indices.

\subsubsection{Exotic matter}
\label{ssubsec:exotic}

To get 
SM chiral matter but not other representations $R'$ of $\gsm$, in the above
procedure, it is important to choose the right embedding. For
directly tuned $\gsm$, it has been argued in \cite{TaylorTurnerGeneric} that the
model generically contains the SM matter fields and the representations
$\left(\mathbf{3},\mathbf{1}\right)_{-4/3},\left(\mathbf{1},\mathbf{2}\right)_{3/2},\left(\mathbf{1},\mathbf{1}\right)_{2}$,
while constructing representations $R'$ other than these (defined as
 exotic matter representations) requires extensive amounts of fine-tuning. Here
we will see that this is no longer the situation in the case of vertical
flux breaking. As described in \S\ref{subsec:breaking}, we
can get exotic $\U(1)$'s from simple vertical flux
constraints, leading
to many possible exotic representations $R'$. Below we give such an example. Note that
when $R'$ other than SM matter representations
are involved, the flux breaking may not realize all
independent sets of chiral matter allowed by anomaly cancellation.

The directly tuned $\gsm$ models containing generic matter, which
includes the Standard Model representations, can be naturally unHiggsed into
$\SU(4)\times \SU(3)\times \SU(2)$ models \cite{Raghuram:2019efb}. It is interesting
that the converse cannot be achieved from the perspective of
$E_7$ flux breaking. As an example, we consider a flux breaking
pattern from vertical flux that can be associated with the breaking
route $E_{7}\rightarrow \SU\left(4\right)\times
\SU\left(3\right)\times \SU\left(2\right)\rightarrow
\SU\left(3\right)^{2}\times
U\left(1\right)/\mathbb{Z}_{3}\rightarrow\gsm$.
Note that in this example we do not use remainder flux; all the
breaking comes from vertical fluxes.
This time we put the non-abelian part in nodes $1,4,5$, so
we turn on nonzero $\Theta_{i'\alpha}$ for
$i'=2,3,6,7$. Then we find that the $\U(1)$ charge is
%\begin{equation}
%    q=\frac{3}{2}b_{2}+\frac{1}{3}b_{3}-\frac{4}{3}b_{6}-2b_{7}\,,
%\end{equation}
given by the generator
\begin{equation}
    T=\frac{1}{2}T_{1}+T_{2}-\frac{1}{3}T_{4}-\frac{2}{3}T_{5}-T_{6}-T_{7}\,.
\end{equation}
Therefore, we further impose $\Theta_{2\alpha}=\Theta_{6\alpha}+\Theta_{7\alpha}$ for all
$\alpha$. This condition does not coincide with any root of
$E_7$, so it really induces an exotic $\U(1)$.

As above, we analyze the breaking of $\mathbf{56}$ and $\mathbf{133}$
separately. The first observation is that $\mathbf{56}$ does not break
into the generic matter representations that appear in directly tuned $\gsm$ models. Instead, it
breaks into the representations
\begin{equation} \label{E7to432fund}
    \left(\mathbf{3},\mathbf{2}\right)_{1/6},\left(\mathbf{3},\mathbf{1}\right)_{5/3},\left(\mathbf{3},\mathbf{1}\right)_{-1/3},\left(\mathbf{1},\mathbf{2}\right)_{1/2},\left(\mathbf{1},\mathbf{1}\right)_{1},\left(\mathbf{3},\mathbf{1}\right)_{-4/3},\left(\mathbf{1},\mathbf{2}\right)_{3/2},\left(\mathbf{1},\mathbf{1}\right)_{2}\,.
\end{equation}
That is, the right-handed up quark is replaced by the exotic
$\left(\mathbf{3},\mathbf{1}\right)_{5/3}$, and there are various
exotic representations that appear. There are three independent sets of chiral matter from anomaly cancellation. The chiral indices from flux breaking, however, only realize two of them:
\begin{equation}
    \chi^{\mathbf{56}}=\Theta^{\mathbf{56}}_{3}\left(1,0,-3,2,-3,1,-1,2\right)+\frac{1}{2}\left(\Theta^{\mathbf{56}}_{6}+\Theta^{\mathbf{56}}_{7}\right)\left(3,-1,-6,3,-7,1,-2,5\right)\,.
\end{equation}
Here the components of the vectors of fields correspond to the matter representations in Eq.\ (\ref{E7to432fund}) in the same order. The first set of
of anomaly-canceling
chiral matter fields only contains the generic matter representations
that appear in directly tuned
$\gsm$ models, while the second set involves the exotic
$\left(\mathbf{3},\mathbf{1}\right)_{5/3}$.

The analysis for $\mathbf{133}$ is  similar. Under the prescribed
breaking route it breaks into
\begin{gather}
    \left(\mathbf{3},\mathbf{2}\right)_{1/6},\left(\mathbf{3},\mathbf{2}\right)_{7/6},\left(\mathbf{3},\mathbf{2}\right)_{-11/6},\left(\mathbf{3},\mathbf{1}\right)_{2/3},\left(\mathbf{3},\mathbf{1}\right)_{-1/3},\left(\mathbf{3},\mathbf{1}\right)_{-4/3},\nonumber\\
    \left(\mathbf{3},\mathbf{1}\right)_{5/3},\left(\mathbf{1},\mathbf{2}\right)_{1/2}, \left(\mathbf{1},\mathbf{2}\right)_{3/2},\left(\mathbf{1},\mathbf{1}\right)_{1},\left(\mathbf{1},\mathbf{1}\right)_{2},\left(\mathbf{1},\mathbf{1}\right)_{3}\,.
\end{gather}
This gives many more exotic matter representations than $\mathbf{56}$,
which can have nontrivial chiral indices for generic fluxes.
Again, only two of
the allowed independent sets of chiral matter are realized.
The chiral indices are
\begin{align}
    \chi^{\mathbf{133}}=&\Theta^{\mathbf{133}}_{3}\left(-1,0,0,0,0,1,1,1,0,1,-1,0\right)\nonumber\\
    &+\left(\Theta^{\mathbf{133}}_{3}+\Theta^{\mathbf{133}}_{6}+\Theta^{\mathbf{133}}_{7}\right)\left(0,2,1,-3,3,-4,2,-6,1,4,-2,1\right)\,,
\end{align}
following the above order. The first set contains only those $R'$ from
both $\mathbf{56}$ and $\mathbf{133}$.
Note that while the $\mathbf{56}$ alone does not generate all states
in the Standard Model spectrum, the missing states are supplied by the
$\mathbf{133}$.
On the other hand, there is no choice of fluxes that gives only SM
matter and no exotics: the second set of representations from
$\mathbf{133}$ is the only place that some exotic matter fields like
$\left(\mathbf{1},\mathbf{1}\right)_{3}$ appear, so the fluxes
generating this family must cancel in the absence of exotic matter.
But this is the only combination that includes the field
$\left(\mathbf{3},\mathbf{1}\right)_{2/3}$, so we cannot get the full
Standard Model chiral matter spectrum from this construction without
at least some exotics.

In conclusion,
while the flux constraints are equally simple in many constructions,
choosing  the incorrect embedding of $\gsm$ into $E_7$ can lead to a
variety of exotic chiral matter fields. These matter representations
are generically present since there are many more than 4
choices of $\U(1)$ in $\gsm$ that can be embedded into $E_7$.
For many of these choices, unlike the case just analyzed here, there
may actually be no resulting fields in some of the Standard Model
representations.  In others, like this one, all of the SM
representations may appear along with some exotics; while in this
specific case we can show that no flux combination is possible that
gives just SM matter without exotics, it is possible that for other
U(1) choices, a judicious tuning of fluxes may cancel the chiral
multiplicities of all exotic matter fields, still allowing for an SM
construction with the expected matter fields and no exotics, but we
leave a full consideration of this question for further research.
%In principle, one can count
%all such embeddings and study the matter spectrum given by
%each embedding. Although the analysis can be tedious, the
%genericity between different embeddings remains an
%interesting open question.

\subsection{Solving the flux constraints}
\label{subsec:solve}

Now we turn back to the construction of  the
Standard Model gauge group and
chiral matter from \S\ref{ssubsec:SMchiral}. Although we have obtained the chiral index in
terms of $\Theta_{i'\alpha}$, we still need to solve the flux
constraints and express everything in terms of flux
parameters $\phi$.

There is a subtlety before breaking $E_7$. Most $E_7$
models have codimension-3 singularities with degree
$(4,6,12)$. Such singularities can no longer be simply
interpreted as Yukawa couplings. The fiber becomes non-flat
at these points and supports an extra vertical flux.
It also seems to correspond to extra strongly coupled
(chiral) degrees of freedom, possibly M5-branes wrapping
non-flat fibers \cite{Candelas:2000nc,Lawrie:2012gg,Jefferson:2021bid}. Although $E_7$ itself does not support any
chiral matter, after flux breaking the extra flux may induce
more chiral matter which is not covered by our formalism.
This will be studied in \cite{46}. For realistic SM-like models, we
simply set such extra flux to vanish, so all the flux we
consider is for flux breaking.

It is now straightforward to solve the flux constraints by
considering independent $S_{i\alpha}$. Recall
from (\ref{eq:theta-matrix}) that
$\Theta_{i\alpha}=-\Sigma\cdot D_\alpha\cdot D_\beta \kappa^{ij} \phi_{j\beta}$.
For independent $S_{i\alpha}$,
the triple intersection form $M^B_{\alpha\beta}=\Sigma\cdot D_\alpha\cdot D_\beta$ on $B$ is invertible. The solution to
$\Theta_{1\alpha}=\Theta_{2\alpha}=\Theta_{3\alpha}=\Theta_{7\alpha}=0$ is simply
%\begin{gather}
%    \phi_{1\alpha}=2n_\alpha\,,\quad\phi_{2\alpha}=4n_\alpha\,,\quad\phi_{3\alpha}=6n_\alpha\,,\nonumber\\
%    \phi_{4\alpha}=5n_\alpha\,,\quad\phi_{5\alpha}=\phi_{5\alpha}\,,\quad\phi_{6\alpha}=\phi_{6\alpha}\,,\quad\phi_{7\alpha}=3n_\alpha\,,
%\end{gather}
\begin{gather}
    \phi_{1\alpha}=2n_\alpha\,,\quad\phi_{2\alpha}=4n_\alpha\,,\quad\phi_{3\alpha}=6n_\alpha\,,\quad
    \phi_{4\alpha}=5n_\alpha\,,\quad\phi_{7\alpha}=3n_\alpha\,,
    \label{eq:phi-n}
\end{gather}
with $\phi_{5 \alpha}, \phi_{6 \alpha}$ arbitrary,
but we pick sufficiently generic
$\phi_{5 \alpha},\phi_{6 \alpha}$ such that the
resulting gauge group
does not get further enhanced.
These fluxes give
\begin{equation}
    \Theta_{4\alpha}=M^B_{\alpha\beta}(\phi_{5\beta}-4n_\beta)\,,\quad\Theta_{5\alpha}=M^B_{\alpha\beta}(5n_\beta-2\phi_{5\beta}+\phi_{6\beta})\,,\quad\Theta_{6\alpha}=M^B_{\alpha\beta}(\phi_{5\beta}-2\phi_{6\beta})\,.
\end{equation}
%\drop{The
%condition that the ratio
%$\phi_{i\alpha}/\phi_{i\alpha'}$ is independent of $i$ is
%satisfied as long as the  components $n_\alpha,\phi_{5\alpha},\phi_{6\alpha}$
%satisfy it separately.} 
The flux quantization condition is satisfied by integer $\phi_{i\alpha}$, hence integer $n_\alpha$ when $c_2(\hat Y)$ is even. The D3-tadpole condition is satisfied when $\phi_{i\alpha}$ are sufficiently small. Now Eq.\ (\ref{chi56}) and (\ref{chi133}) give
\begin{equation} \label{generalGSMchi}
    \chi_{(\mathbf 3,\mathbf 2)_{1/6}}=\Sigma\cdot(6K_B+5\Sigma)\cdot D_\alpha n_\alpha\,.
\end{equation}
This is one of the main results in this paper.
The independence of the chiral multiplicity from the parameters $\phi_{5 \alpha}, \phi_{6 \alpha}$ can be understood from the fact that these fluxes do not hit the roots of the preserved part of the gauge group.
Note that
$-(6K_B+5\Sigma)$ is the class of the coefficient of $s^5 z^6$ in the $E_7$ Tate model \cite{BershadskyEtAlSingularities,KatzEtAlTate}. Intersecting it with
$C_{\mathbf{56}}$ gives the codimension-3 singularities.

We see that in a generic basis for the base divisors $D_\alpha$, there are $r$ (see \S\ref{subsec:breaking}) quantized flux parameters 
contributing to a single
chiral index, and the chiral index has a linear
Diophantine structure. This is unlike the case in directly
tuned $\gsm$ models, where the chiral index is controlled by
a single flux parameter with a large constant factor, and
either specific geometries must be chosen, or a better understanding of the quantization conditions discussed in \S\ref{subsec:flux} must be achieved, to make
the chiral index as small as 3.
In our case, generically the
intersection numbers $\Sigma\cdot(6K_B+5\Sigma)\cdot D_\alpha$ have no common factors, and the chiral index can be
any integer. A generic flux configuration has both positive
and negative $n_\alpha$ with small magnitudes
(due to the large number of flux directions that can contribute to the
tadpole as discussed in \S\ref{subsec:flux}), making the terms in
Eq.\ (\ref{generalGSMchi}) cancel and naturally leading to a small
chiral index. Heuristically, if we sample $\chi_{(\mathbf 3,\mathbf 2)_{1/6}}$ throughout the landscape, we expect a
distribution peaking at $\chi=0$ and decaying as $\chi$
becomes large \cite{Andriolo:2019gcb}.
Therefore, $\chi=3$ is a natural solution although it may not
be the most preferred. In conclusion, Eq.\ (\ref{generalGSMchi}) is favored by phenomenology.

There may be also some rare cases where the triple
intersection numbers have a common factor. Most probably the
common factor forbids the possibility of $\chi=3$, but if the
common factor is 3, interestingly $\chi=3$ becomes both the
minimal and natural nontrivial chiral spectrum.

The appearance of a nontrivial minimal multiplicity, and other aspects
of multiplicity quantization, can be understood in terms of
intersection theory on the base $B$, combined with the structure of
the $E_7$ lattice.  The intersection product $C = \Sigma \cdot (6 K_B +5 \Sigma)$
is a curve in integer homology of the base $B$.  For generic choices of characteristic
data, we expect that this curve will be primitive, in which case
Poincar\'{e} duality asserts that there is a divisor $D' = D_\alpha
n'_\alpha$ with $C \cdot D' = 1, n'_\alpha \in\Z$.  This is the generic
case described above where there are no common factors and the chiral
index can be any integer.  Thus, in some sense the flux associated
with the chiral index can be characterized by a single parameter
$\lambda$, with $n_\alpha = \lambda n'_\alpha$.  On the other hand, a
full treatment of the proper basis for fluxes would involve
identifying flux directions with minimal tadpole contribution, which
we do not investigate further here.
When $C = m C'$ is not primitive but is an integer multiple of a
primitive curve $C'$, this corresponds to the situation where there
are common factors and there is a non-unit minimal multiplicity for
$\chi$; the case where $\chi = 3$ is minimal corresponds to the
situation where $m = 3$.

It is interesting that more
generally we can consider fluxes in $H^4 (\hat{Y},\Z)$ that
may lead to fractional values of $n_\alpha$.
Since  $H^4 (\hat{Y},\Z)$ has a unimodular intersection form, we
expect that there may be fluxes in $H^4 (\hat{Y},\Z)$ with fractional vertical components $n_\alpha$,
when the fluxes $\phi_{i \alpha}$ lie
in the dual of the root lattice (i.e., the weight lattice) of $E_7$.
From the form of the
$E_7$ lattice, the only non-integer fluxes allowed have half-integer
entries for $\phi_{4 \alpha}, \phi_{6 \alpha}, \phi_{7 \alpha}$.
This results in half-integer $n_\alpha$ in Eq.\ (\ref{eq:phi-n}) and
naively appears to lead to half-integer
multiplicities in Eq.\ (\ref{generalGSMchi}). The multiplicities should be, however, guaranteed to be integers from the structure of $H^4 (\hat{Y},\Z)$.
%Thus, either such
%situations are not possible, or the necessary horizontal flux that
%must be included to give an integer flux vector stabilize moduli to
%change the geometry, or something else fixes up the physics of these
The explicit form of $H^4 (\hat{Y},\Z)$ is not yet fully elucidated,
as discussed in  \S\ref{subsec:flux}, which makes these issues a bit
subtle,
and we leave a more detailed investigation of such situations
to further work.

There are still other flux constraints remaining such as
primitivity. Solving these constraints will be demonstrated in Section \ref{sec:example}.

\section{Breaking other gauge groups}
\label{sec:othergroups}

So far we have focused on the breaking $E_7\rightarrow\gsm$,
while among the rigid gauge groups, $E_6$ and $E_8$ can also
be broken into $\gsm$. $E_6$ has been one of the traditional
GUT groups, so its breaking is less novel than $E_7$'s. 
Flux breaking of (non-rigid) $E_6$ F-theory models has been described
in the dual heterotic framework in \cite{Chen:2010tg}.
On
the other hand, $E_6$ also appears in a significant portion of the
landscape, and we therefore generalize our construction to $E_6$
for completeness. $E_8$ is clearly the most abundant
exceptional gauge
group in the landscape, but unfortunately our formalism does
not work for $E_8$ for several reasons. Below we discuss
these two gauge groups separately.

\subsection{$E_6$}
\label{subsec:e6}

It is clear that the above construction for $E_7$ also works
for $E_6$, since we can first break $E_7$ down to $E_6$ in
our breaking by vertical flux. The generalization to $E_6$, however, is more
nontrivial since $E_6$ itself supports chiral matter without
flux breaking. There are more flux parameters $\phi_{ij}$ to
turn on, and both terms in Eq.\ (\ref{fullchiR'}) contribute
to the chiral indices. Although as discussed before the two
terms in Eq.\ (\ref{fullchiR'}) are independent contributions
controlled by different flux parameters, the flux
configuration itself becomes more nontrivial due to flux
quantization. This will be explained below.

\begin{figure}[t]
\centering
\includegraphics[width=0.4\columnwidth]{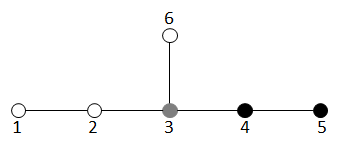}
\caption{The Dynkin diagram of $E_6$. The Dynkin node labelled
$i$ corresponds to the exceptional divisor $D_i$. The solid
nodes are the ones we break to get the Standard Model gauge
group and chiral matter. Node 3  (gray) is broken by
remainder flux while the others are broken by vertical flux.}
\label{dynkine6}
\end{figure}

Although we expect the
middle intersection form on vertical fluxes (\ref{eq:mr-matter}) as
well as the
physics to be resolution-independent,
in practice it is useful to work with a certain resolution,
and extract resolution-independent information from the
results. Here we choose the resolution studied in
\cite{Esole:2017kyr,Bhardwaj_2019,Jefferson:2021bid},  which
for completeness we review in Appendix \ref{sec:resolution}. The exceptional
divisors and the broken directions are described as in Figure \ref{dynkine6}.
The result in \cite{Jefferson:2021bid} shows that when the gauge group is
unbroken, there is a chiral index for the $E_6$ fundamental
$\mathbf{27}$, given by
\begin{equation}
    \chi_{\mathbf{27}}=\Theta_{24}\,.
\end{equation}
Following the procedure in Section \ref{subsec:matter}, we find that the
matter surface $S_0(\mathbf{27})$ is
\begin{equation}
    S_0(\mathbf{27})=D_2\cdot
    D_4+\frac{1}{3}\pi^*(3K_B+2\Sigma)\cdot(-D_1+D_2+2D_4+D_5)\,,
\label{eq:e6-s0}
\end{equation}
where the class $(3K_B+2\Sigma)$ appears in the matter curve
$C_{\mathbf{27}}=-\Sigma\cdot(3K_B+2\Sigma)$.

Recall that both the flux configuration and chiral indices
can be separated into parts that preserve or break the gauge
group. We now decompose $G_4=G_4^{p}+G_4^{b}$, where $G_4^p$
lives in the directions in $M_{\mathrm{phys}}$ and preserves
the gauge group, while $G_4^b$ is the gauge-breaking flux.
Correspondingly we define the flux parameters $\phi^p,\phi^b$
and chiral indices $\chi^p,\chi^b$. By solving the flux
constraints, $G_4^p$ is given by
\begin{equation}
    G_4^p=\phi_{24}^p\left[S_0(\mathbf{27})\right]\,,
\end{equation}
inducing
\begin{equation}
\label{e6chib4breaking}
    \chi^p_{(\mathbf 3,\mathbf 2)_{1/6}}=\chi_{\mathbf{27}}=\frac{1}{3}\Sigma\cdot(3K_B+2\Sigma)\cdot(6K_B+5\Sigma)\phi_{24}^p\,.
\end{equation}
As expected, this chiral index is controlled by a single flux
parameter $\phi^p_{24}$.  We discuss the detailed quantization
condition on this parameter below, but note that $K_B \cdot \Sigma
\cdot \Sigma$ is always even, by the Hirzebruch-Riemann-Roch theorem
for surfaces, so the chiral multiplicity is always an integer when
$\phi_{24}^p$ is a multiple of 3/2.

Now we turn to $G_4^b$. Here we follow the breaking route
used in Section \ref{subsec:solve}. In principle, we can
apply the same procedure as in Section \ref{sec:SM} to
obtain $G_4^b$ and $\chi^b$, but there is a faster way using
the result for $E_7$. We can first break $E_7$ down to
$E_6$ by removing node 6 in Figure \ref{dynkine7}. Note that
$\phi_{6\alpha}$ in the $E_7$ model are completely
independent of other flux parameters and do not contribute to
the chiral indices. Therefore, $G_4^b$ for
$E_6$ is the same as that for $E_7$ by ignoring node 6 in
$E_7$. That is
\begin{gather}
    \phi^b_{1\alpha}=2n_\alpha\,,\quad\phi^b_{2\alpha}=4n_\alpha\,,\quad\phi^b_{3\alpha}=6n_\alpha\,,\nonumber\\
    \phi^b_{4\alpha}=5n_\alpha\,,
    %\quad\phi_{5\alpha}=\phi_{5\alpha}\,,
    \quad\phi^b_{6\alpha}=3n_\alpha\,,
    \label{eq:e6fluxbreaking}
\end{gather}
where $\phi^b_{5 \alpha}$ is arbitrary.
Therefore, the chiral index from flux breaking is
\begin{equation}
\label{e6chibreaking}
    \chi^b_{(\mathbf 3,\mathbf 2)_{1/6}}=\Sigma\cdot(6K_B+5\Sigma)\cdot D_\alpha n_\alpha\,,
\end{equation}
which is exactly the same as Eq.\ (\ref{generalGSMchi}). The
total chiral index is then $\chi=\chi^p+\chi^b$. Despite the
extra $\chi^p$, with the inclusion of $\chi^b$ it is
qualitatively the same as that in $E_7$ models.

So far $\phi^p$ and $\phi^b$ are totally separated, but it
becomes more interesting when flux quantization is
considered. First, in $E_6$ models it is unavoidable to have
a non-even $c_2(\hat Y)$. Using the techniques in \cite{Jefferson:2021bid}, we find
that for our choice of resolution (in terms of independent surfaces)
\begin{align}
    \left[c_2(\hat Y)\right]=&\left[c_2(B)\right]+11\pi^{*}K_{B}^{2}+\left(-12D_{0}+17D_{1}+27D_{2}+30D_{3}+24D_{4}+11D_{5}+14D_{6}\right)\cdot\pi^{*}K_{B} \nonumber \\
    &+\left(6D_{1}+8D_{2}+6D_{3}+6D_{4}+2D_{5}+2D_{6}\right)\cdot\pi^{*}\Sigma+D_2\cdot D_4\,.
\end{align}
Note that $\left[c_2(B)\right]+\pi^{*}K_{B}^{2}$ is always
even \cite{Collinucci:2010gz}. 
%If the gauge group is unbroken i.e. $G_4^b=0$, we
%see that flux quantization generically requires 
%$\phi_{24}^p=3(2k+1)/2$
%for some integer $k$. The situation, however, is different if
%both $G_4^p$ and $G_4^b$ are present. The crucial point is
%that flux quantization only applies to the total flux $G_4$,
%allowing more flux configurations if we look at one of the
%sectors only. In particular, now $\phi_{24}^p$ can be any
%half-integer, as long as appropriate \emph{fractional}
%$\phi_{i\alpha}$ are turned on such that the total flux is
%correctly quantized. 
Note also from the form of Eq.\ (\ref{eq:e6-s0}) that the part of $c_2$
that is odd precisely contributes to a half-integer contribution to $G_4^p$
 and does not necessitate breaking of the $E_6$.
If the gauge group is unbroken i.e. $G_4^b=0$, we
see that flux quantization generically requires 
$\phi_{24}^p=3(2k+1)/2$
for some integer $k$. 
(As noted above, this always gives integer chiral multiplicity since  $K_B \cdot \Sigma \cdot \Sigma$ is
always even.)
The situation, however, is different if
both $G_4^p$ and $G_4^b$ are present. The crucial point is
that flux quantization only applies to the total flux $G_4$,
allowing more flux configurations if we look at one of the
sectors only. In particular, now $\phi_{24}^p$ can be any
half-integer, as long as appropriate \emph{fractional}
$\phi_{i\alpha}$ are turned on such that the total flux is
correctly quantized. Therefore, the presence of
gauge-breaking flux enlarges the possibilities of matter
surface flux, although they contribute to chiral indices
independently.

\subsection{$E_8$}
\label{subsec:e8}

It is tempting to apply our formalism to $E_8$ models.
Nevertheless, these models have very different physics from
$E_7,E_6$ models, and the direct construction of Standard Model-like
vacua from flux breaking of rigid $E_8$ factors fails for various
reasons.

The first reason is that an $E_8$ model generically contains
codimension-2 $(4,6)$ singularities. While this type of
singularity in 4D has not been completely understood, it is
believed to be parallel to the story in 6D F-theory models.
There, a simple physical interpretation of these singularities is
obtained by blowing up the locus into the tensor branch till
the singularities are within the minimality bound i.e. degree
$<(4,6,12)$. The origin of  the tensor branch corresponds to
shrinking the resulting exceptional divisors to zero volume,
giving a strongly coupled limit of the model. D3-branes
wrapping these exceptional divisors also become tensionless
strings in the low-energy theory. All these signal the
presence of strongly coupled superconformal sectors
\cite{HeckmanMorrisonVafa,DelZotto:2014hpa,Apruzzi:2018oge}. These
extra degrees of freedom, called conformal matter, are not
covered by our formalism for analyzing flux breaking.

Still, there are $E_8$ models without these kinds of
singularities and naively our formalism should work in such cases.
 The second reason, however, that these geometries are problematic is
 that the condition for the absence of
conformal matter is that the codimension-2 singularity has
trivial homology class i.e. $\Sigma\cdot(6K_B+5\Sigma)=0$.
Surprisingly, using Eq.\ (\ref{generalGSMchi}) this immediately
implies that no chiral matter can be induced, even if we
break $E_8\rightarrow\gsm$. It remains interesting to find a
reason behind this apart from direct computations.

All this seem to suggest that the class of SM-like models we
have constructed may still not be the largest class in the
landscape. In particular, the F-theory geometry with the most
flux vacua contains many factors of $E_8$, but
no factors of $E_7,E_6$ and does not
support our formalism \cite{TaylorWangVacua}. In principle, the most generic SM
matter should come from the strongly coupled matter in $E_8$.
Some initial investigation into studying the 4D spectrum from strongly
coupled $E_8$ matter
in the context of E-string
theory is described in \cite{TianWangEString}.

\section{An explicit example}
\label{sec:example}

The above construction of SM gauge group and chiral matter can be done on a large class of bases
containing rigid or tuned $E_7,E_6$ factors. For the rigid case, we need
a non-toric base such that there can be a rigid divisor
supporting hypercharge flux. In this section, we
provide an explicit example of such a construction,
with three generations of SM chiral matter as the
minimal and preferred chiral matter content. Since
the $E_7$ models require a more involved
construction with $r\geq 4$, below we construct
rigid $E_6$ with $r=3$. As shown below, the gauge divisor is a del
Pezzo surface $dP_4$, so the model has a limit where gravity is
fully decoupled \cite{BeasleyHeckmanVafaII}.

We choose a base $B$ through the following procedure: first consider
$A=\mathbb P^1\times \mathbb F_1$ where $\mathbb F_1$ is the Hirzebruch
surface. Then $B$ is a certain hypersurface in an
ambient space $X$ that is a $\mathbb P^1$-bundle over $A$
with a certain normal bundle. This example is
a generalization of an example of a geometry supporting remainder flux in \cite{Braun:2014pva}; more
generally we can similarly analyze any hypersurface $B$ in a toric
fourfold $X$ as long as $B$ is ample.
In the explicit example we
consider here,
both vertical and remainder fluxes are
incorporated, and all flux constraints can be explicitly
solved.

Let us first construct the ambient space $X$. To
construct a model with $r\geq 3$ (see
\S\ref{subsec:breaking}), we need to start with a threefold
$A$ with $h^{1,1}\geq 3$.
 As an example with $h^{1,1}(A)=3$, we choose $A$ to be
$\mathbb P^1\times \mathbb F_1$. Within $\mathbb F_1$,
we denote $s$ as the
$\mathbb P^1$ section and $f$ as the $\mathbb P^1$ fiber. Then
the intersection numbers are $f^{2}=0,f\cdot s=1,s^2=-1$.
Now on $A$, we denote $\sigma$ as the
$\mathbb{F}_{1}$ section and $S,F$ as the
$\mathbb P^1$ product with $s$ and $f$ respectively. Then
the anticanonical class of $A$ is
$-K_{A}=2\sigma+2S+3F$.
\begin{comment}
We denote $\sigma$ as the $dP_2$ section and
$S_1,S_2,H$ as the $\mathbb P^1$ product with the
two exceptional divisors and the hyperplane of
$dP_2$ respectively. Note that the effective
divisors are spanned by $\sigma,S_1,S_2$, and
$F=H-S_1-S_2$. The anticanonical class of $A$ is $-K_A=2\sigma+2S_1+2S_2+3F$.
\end{comment}
The nonzero intersection numbers are:
\begin{equation} \label{p1crossdp2}
    \sigma\cdot S\cdot F=1\,,\;\sigma\cdot S^2=-1\,.
\end{equation}
Finally we can
describe $X$ as a $\mathbb P^1$-bundle over $A$. We denote
$\sigma_A$ as the section and $F_\sigma,F_S,F_F$ as the
fibers along $\sigma,S,F$ respectively. Let the normal bundle
be $N_A=-a\sigma-bS-cF$ where $a,b,c\in\mathbb Z_{\geq0}$.
Then its anticanonical class is
$-K_X=2\sigma_A+(a+2)F_\sigma+(b+2)F_S+(c+3)F_F$. The
intersection numbers can be calculated using Eq.\ (\ref{p1crossdp2}) and the relations
$\sigma_A\cdot(\sigma_A+aF_\sigma+bF_S+cF_F)=0$. Note that with the below choice of $N_A$, $X$ is a
smooth, projective toric variety with a unique triangulation.

We then choose the base as a hypersurface in $X$ with irreducible class
$B=\sigma_A+(a+1)F_\sigma+(b+1)F_S+(c+2)F_F$. By abuse of notation, we use
$B$ to denote both the base and its divisor class in $X$. By
adjunction we have $-K_B=B\cdot(\sigma_A+F_\sigma+F_S+F_F)$.
As shown below, $B$ is strictly inside the K\"ahler cone of $X$, so $B$ is ample in $X$.  By Lefschetz's hyperplane theorem, we then have the isomorphism
$H^{1,1}\left(B,\mathbb{Z}\right)\cong H^{1,1}\left(X,\mathbb{Z}\right)$. In other words, divisors
on $B$ are spanned by the intersections $B\cdot\sigma_A,B\cdot F_\sigma,B\cdot F_{S},B\cdot F_{F}$.
The intersection numbers relevant to our purpose are
\begin{gather}
    B\cdot\sigma_A\cdot F_\sigma^2=0\,,\;B\cdot\sigma_A\cdot F_\sigma\cdot F_S=1\,,\;B\cdot\sigma_A\cdot F_\sigma\cdot F_F=1\nonumber \\
    B\cdot\sigma_A\cdot F_S\cdot F_F=1\,,\;B\cdot\sigma_A\cdot F_S^2=-1\,,\;B\cdot\sigma_A\cdot F_F^2=0\,.
    \label{Sigmaintersections}
\end{gather}

\begin{comment}
By choosing $N_{\Sigma}=-8S-7F$, there is a rigid $E_7$
supported on $\Sigma$ (see Section \ref{subsec:groups}) and an even $c_2(\hat Y)$. The nonzero intersection numbers are then
$\Sigma\cdot F_{S}\cdot F_{F}=1,\Sigma^{2}\cdot F_{F}=-8,\Sigma\cdot F_{S}^{2}=-1,\Sigma^{2}\cdot F_{S}=1,\Sigma^{3}=48$.
\end{comment}

Now consider the gauge divisor
$\Sigma=B\cdot\sigma_A=\sigma_A\cdot(F_\sigma+F_S+2F_F)$.
To determine the rigid gauge group on $\Sigma$, we
calculate
\begin{gather}
    -K_{\Sigma}=B\cdot\sigma_A\cdot \left(F_\sigma+F_S+F_F\right)=\sigma_A\cdot\left(2 F_\sigma\cdot F_S+3F_\sigma\cdot F_F+2F_S\cdot F_F\right)\,, \\
    N_{\Sigma}=B\cdot\sigma_A^2
    =-\sigma_A\cdot\left((a+b)F_\sigma\cdot F_S+(2a+c)F_\sigma\cdot F_{F}+(b+c)F_S\cdot F_F\right)\,.
\end{gather}
By the conditions in
\S\ref{subsec:groups}, we choose $(a,b,c)=(3,3,3)$ such that $\Sigma$
is a rigid divisor
supporting a rigid $E_6$. Note that with this choice of
$N_A$, $C_\mathbf{27}$ is trivial and all
the matter is in the  $E_6$ adjoint $\mathbf{78}$ before flux breaking.
Note also that with this choice $N_\Sigma = 3K_\Sigma$ is divisible by 3, so the curve $C =\Sigma \cdot
(6K_B + 5 \Sigma)$ appearing in Eq.\ (\ref{e6chibreaking}) is not primitive
but takes the form $C = 3C'$ as discussed in \S\ref{subsec:solve}, and
we expect chiral multiplicities that are multiples of 3.
Note further that we have not ruled out fully the possibility of increased
enhancement over curves in the base, which might in principle give
rise to additional surfaces.  Even if this occurs,
however, it should not be relevant for our construction as we can
simply keep any associated additional fluxes that may arise to
vanish.  For future work and more general constructions, however, it
would be useful to develop more completely the methodology for
analyzing the structure of hypersurface bases of this general kind.

We can solve the flux constraints after determining the
geometry. First we focus on vertical flux. Let us analyze the constraint from
primitivity. The
independent $S_{i\alpha}$ are $S_{i(B\cdot F_\sigma)},S_{i(B\cdot F_S)},S_{i(B\cdot F_F)}$, while $S_{i\Sigma}$ is a linear
combination of the former three. Using the
intersection numbers, we see that the triple intersection
form $M^B_{\alpha\beta}$ has one positive and two negative
directions, so primitivity can be satisfied.
As explained in Section \ref{subsec:primitive}, we focus on
the K\"ahler form of the base $J_B$, which can be expanded using a
basis of base divisors (recall that in this case as noted above,
$H^{1,1}\left(B,\mathbb{Z}\right)\cong
H^{1,1}\left(X,\mathbb{Z}\right)$):
\begin{equation}
    \left[J_{B}\right]=B\cdot\left(t_{1}F_F+t_{2}\left(F_S+F_{F}\right)+t_{3}F_\sigma+t_{4}\left(\sigma_A+3F_\sigma+3F_S+3F_F\right)\right)\,,
\end{equation}
where $t_{1},t_{2},t_{3},t_{4}$ are linear
combinations of K\"ahler moduli,
%on $B$
 and may be negative
inside the K\"ahler cone of $B$
in general. While determining the
exact K\"ahler
cone of a hypersurface in a toric variety can be subtle, the K\"ahler cone of $B$ must contain that of $X$
 \cite{Demirtas:2018akl}.
For simplicity, we look for a solution of the primitivity constraints in
the K\"ahler cone of $X$ only.
First, the
K\"ahler cone of $X$ can be obtained from the Mori cone, which is
spanned by $F_\sigma\cdot F_S\cdot F_F,\sigma_A\cdot F_\sigma\cdot F_S,\sigma_A\cdot F_\sigma\cdot F_F,\sigma_A\cdot F_S\cdot F_F$. By computing the
dual cone, we see that the interior of the K\"ahler
cone of $X$ corresponds to $t_{1},t_{2},t_{3},t_{4}>0$. Now primitivity implies that for all $i$
\begin{equation}
    t_1(\phi_{i\sigma}+\phi_{iS})+t_2(2\phi_{i\sigma}+\phi_{iF})+t_3(\phi_{iS}+\phi_{iF})=0\,.
\end{equation}
To determine the chiral matter spectrum, we first
focus on $n_\alpha$. There must be a pair of
coefficients of $t_a$ with opposite signs, which
places constraints on the possible $n_\alpha$.

\begin{comment}
There is another constraint on $n_\alpha$ from flux
quantization. Interestingly, our example has
a non-even $c_2(\hat Y)$. Using the techniques in \cite{Jefferson:2021bid} and the choice of resolution in Appendix \ref{sec:resolution}, we find
that for models with a single $E_7$: (in terms of homologically independent surfaces)
\begin{align}
    \left[c_2(\hat Y)\right]=&\left[c_2(B)\right]+11\pi^{*}K_{B}^{2}\nonumber \\
    &+\left(-12D_{0}+14D_{1}+30D_{2}+48D_{3}+41D_{4}+28D_{5}+17D_{6}+27D_{7}\right)\cdot\pi^{*}K_{B} \nonumber \\
    &+\left(2D_{1}+6D_{2}+12D_{3}+12D_{4}+8D_{5}+6D_{6}+8D_{7}\right)\cdot\pi^{*}\Sigma\,.
\end{align}
Note that $\left[c_2(B)\right]+\pi^{*}K_{B}^{2}$ is always
even \cite{Collinucci:2010gz}. Since $\Sigma.K_B=K_\Sigma-N_\Sigma$, Eq.\ (\ref{quantization})
\end{comment}

Now we consider the chiral index. First, $\phi^p$ and Eq.\ (\ref{e6chib4breaking}) vanish since $C_\mathbf{27}$ is
trivial. Eq.\ (\ref{e6chibreaking}) then gives
\begin{equation}
    \chi_{\left(\mathbf{3},\mathbf{2}\right)_{1/6}}=-3(2n_\sigma+n_S+2n_F)\,,
\end{equation}
where $n_I=n_{(B\cdot F_I)}$. 
As discussed before, it
is natural to consider small $\phi_{i\alpha}$. To
fulfill flux quantization as in Section \ref{subsec:e6}, we turn on  integer $n_\alpha$.
One of the minimal flux configurations satisfying the flux
constraints has $(n_\sigma,n_S,n_F)=(-1,1,0)$, hence
gives $\chi=3$ as the minimal and preferred chiral spectrum.

We now turn to remainder flux. It can be shown that $\Sigma$ is a del
Pezzo surface $dP_4$ and supports remainder flux. First notice that
$\Sigma$ is a hypersurface in $A$
with class $\sigma+S+2F$. In other words, $\Sigma$ is the vanishing
locus
\begin{equation} \label{Sigmalocus}
    xP+yP'=0\,,
\end{equation}
in $A$, where $P,P'$ are sections of $\mathcal O_A(S+2F)$, and $x,y$
are the homogeneous coordinates of the $\mathbb P^1$ in $A$. For generic
points in the $\mathbb F_1$, Eq. (\ref{Sigmalocus}) has a unique
solution, representing a single point in $\mathbb P^1$. On the other hand,
there are $(s+2f)^2=3$ points in $\mathbb F_1$ such that $P=P'=0$,
and Eq. (\ref{Sigmalocus}) represents the whole $\mathbb P^1$.
Therefore, the geometry of $\Sigma$ is $\mathbb F_1$ blown up in 3
generic points i.e. a $dP_4$, where the projection $A \rightarrow
\F_1$ gives the blow-down map
$\Sigma \rightarrow \F_1$.

To construct the remainder flux, notice that the three exceptional
curves on $\Sigma$ from blowing up $\mathbb F_1$ (denoted by
$e_1,e_2,e_3$) are all $\mathbb P^1$
fibers in $A$, which have class $S\cdot F$. Under the
inclusion map $\iota:\Sigma\rightarrow B$, we then have $\iota_* e_i=\sigma_A\cdot F_S\cdot F_F$
for all $i=1,2,3$.  Therefore, we can choose e.g. $C_{\mathrm{rem}}=e_1-e_2$ and turn on the remainder flux
\begin{equation} \label{eq:e6remainderflux}
    G_4^\mathrm{rem}=\left[\left(D_Y+\phi_{4r}D_4+\phi_{5r}D_5\right)|_{C_\mathrm{rem}}\right]\,,
\end{equation}
where the first term is the hypercharge flux in Eq.\ (\ref{hyperchargeflux}) and the other two terms with free flux
parameters $\phi_{4r},\phi_{5r}$ do not affect the gauge group. Notice
that when $\phi_{4r}=4$, it is known that this flux removes the exotic
vector-like $(\mathbf 3,\mathbf 2)_{-5/6}$ \cite{BeasleyHeckmanVafaII},
avoiding a number of phenomenological inconsistencies such as part of proton decay. Importantly,
this removal can be achieved without the complication of using fractional line bundles,
which must be used in traditional $\SU(5)$ models, due to more flux
parameters. This is parallel to how the Diophantine structure in vertical flux enables much more possibilities for chiral indices.
These further phenomenological features of our models will be studied
in 
a future publication. In the analysis below, we keep this choice of $\phi_{4r}$.

For consistency, we still need to study the tadpole condition, see Eq.\ (\ref{tadpole}). First we specify the remaining flux parameters
$\phi_{5\alpha},\phi_{5r}$. As an example with small tadpole, we choose
$(\phi_{5F_\sigma},\phi_{5F_S},\phi_{5F_F},\phi_{5r})=(-3,2,1,2)$. The
total flux is then given by these four parameters, the vertical flux in
Eq. (\ref{eq:e6fluxbreaking}), and the remainder flux in Eq.
(\ref{eq:e6remainderflux}) with $\phi_{4r}=4$. The
vertical flux consistently stabilizes the K\"ahler moduli at
$2t_1=2t_2=t_3$. The total tadpole is
\begin{equation}
    \frac{1}{2}\int_{\hat Y}G_4^\mathrm{vert}\wedge G_4^\mathrm{vert}+\frac{1}{2}\int_{\hat Y}G_4^\mathrm{rem}\wedge G_4^\mathrm{rem}=13+6=19\,.
\end{equation}
Here we have naturally extended Eq. (\ref{eq:simple-mr}) to remainder
flux, since the intersections are all localized on $\Sigma$. 
Using the
technique in \cite{Esole:2017kyr}, we find that $\chi(\hat Y)=2088$.
Since $\chi(\hat Y)/24=87>19$, the tadpole condition is satisfied.

We would like to emphasize that although we have chosen an
explicit global example here, the analysis is purely local.
The same breaking pattern and matter spectrum are expected
whenever there is a $\Sigma$ in $B$ with the same geometry and normal bundle. This is analogous to 6D
F-theory models, in which any curve of self-intersection $-6$
supports a rigid $E_6$ \cite{MorrisonTaylorClusters}. We expect that many of the F-theory threefold
bases contain the above local structure. Moreover, there are
lots of local structures throughout the landscape that
support the same flux breaking. Therefore, our construction
provides a large class of models with SM gauge group and
chiral matter, with 
much less fine-tuning that is needed for other known constructions.

\section{Conclusion and further questions}
\label{sec:Conclusion}

\subsection{Summary of results}
\label{subsec:summary}

In this paper, we have described a large class of  Standard
Model-like
models with the right gauge group and chiral matter spectrum,
using the framework of F-theory compactifications. These
models originate from rigid $E_7$ (covered briefly in \cite{Li:2021eyn}) or $E_6$ gauge symmetries,
which are ubiquitous in the string landscape and do not
require any fine-tuning of moduli. In particular, the UV
physics of string theory allows us to use $E_7$, in
addition to the traditional $E_6$, as a GUT group. The same
construction can be  carried out on many (but non-toric) F-theory threefold
bases that contain rigid $E_7$ or $E_6$ local structures. Due to
such genericity, we expect that this is a natural way for 
the Standard Model
to arise in the landscape. Although we do not have an exact
quantification, we believe these models should be more
generic than tuned SM-like models in the landscape.

Remarkably, these models also enjoy the advantage of typically having
small chiral indices. While the chiral indices in tuned
SM-like models are usually too large unless very specific geometries are considered, or the subtle flux quantization issues discussed in \S\ref{subsec:flux} are managed,
the chiral indices in our models
have a linear Diophantine structure that naturally leads to
small integers for typical geometries. As a result, three generations of SM chiral
matter can be easily realized in our models. In particular, a
subset of them have $\chi=3$ as the minimal or preferred
matter content. This is favored by phenomenology. We hope
that this large class of SM-like constructions can shed some
light on where our Universe sits in the string landscape, and
whether it is a \emph{natural} solution in the landscape.

The main tool we have used to achieve the above results is gauge
symmetry breaking with both vertical and remainder fluxes. This
is an efficient way to build models, as it breaks the
gauge group and induces chiral matter at the same time. While
this idea is not new, we have developed it here in depth to give a
systematic procedure to describe the flux breaking from any
$G$ to any $G'$ on almost any base, and calculate the chiral
spectrum induced by the vertical flux. All these calculations can be done
using simple formulas and give results that are manifestly
resolution-independent, with the base geometry and group
theory data as the only input. A remarkable fact from this
procedure is that even if $G$ does not support any chiral
matter, generically a chiral spectrum is still induced if $G'$
supports chiral matter. This is why we can use $G=E_7$ in our
SM-like models. The only exception we find is
$G=E_8$. The procedure developed here is a byproduct of our study of SM-like
models, and should be useful for other types of F-theory
model building in the future.

\subsection{Further questions}
\label{subsec:questions}

As mentioned at the beginning of the paper, although the models we have
constructed here
have the right gauge group and chiral matter spectrum, they
are far from complete in realizing the full details of the
 Standard Model in string theory. 
More work is needed to understand the full matter spectrum including
vector-like fields, Yukawa couplings, questions related to proton
decay, etc.
Many other more general
questions can also be asked, regarding
both theoretical and phenomenological aspects of these models. Examples include:

\begin{itemize}
    \item One interesting feature of our formalism of flux
    breaking is that it intrinsically relies on the
    non-perturbative physics of F-theory. The gauge-breaking
    flux we turn on does not have any immediately obvious
    description in the low-energy theory. In particular, the
    approach of inducing chiral matter with the flux cannot be
    realized in the framework of field theory in any known way. Although the
    broken gauge group and chiral matter spectrum are
    certainly low-energy observables, they do not give full
    information on the flux configuration, and the original $E_7$ or
    $E_6$ gauge group does not seem to be apparent in any clear way in
    the low-energy theory. To gain a more
    complete picture, it would be interesting to understand the
    structure of these models better from the low-energy perspective
    and/or in the dual heterotic  framework.
    
    \item In a string compactification compatible with observations,
      the moduli must be stabilized. In F-theory, the stabilization of
      complex structure moduli is done by turning on horizontal flux,
      inducing a superpotential for the moduli. This flux is
      orthogonal to vertical flux and does not affect the matter
      spectrum. In models with tuned gauge groups, however, some
      complex structure moduli must be fixed and this complicates the
      problem of computing the period vectors, hence superpotential,
      when combined with these tunings. On the other hand, our models rely on
      rigid gauge groups and there is no constraint on complex
      structure moduli. Therefore, the stabilization can be done
      independently without affecting the gauge sectors.  This
      promises, in principle, to make the calculation of moduli
      stabilization easier,
      and opens up an interesting possibility of finding SM-like
      models with moduli stabilized, along the lines of
      \cite{Demirtas:2021nlu,Demirtas:2021ote} and related
      work.\footnote{We thank Manki Kim for discussion on this.}
    
    \item Our construction of SM-like models is
      base-independent.
It is thus possible to apply our
    construction to a large number of explicit F-theory
    threefold bases and perform statistical analysis. 
 There are several distinct such statistical problems of interest.  On
 the one hand, for a given local geometry that supports this
 construction, it will be useful to know  what portion of
 flux configurations can break the rigid gauge group
    down to $\gsm$, and/or give three generations of SM
    chiral matter. 
At the same time it would be desirable to have a better understanding
of the global space of threefold bases that support 4D elliptic
Calabi-Yau spaces, and how ubiquitous the presence of rigid $E_6$ or
$E_7$ gauge factors is in this space.
In particular,
while the current list of F-theory threefold
    bases is far from complete, the large ensembles of toric bases
    considered in
    \cite{TaylorWangMC,HalversonLongSungAlg,TaylorWangLandscape}
suggest that $E_6$ and $E_7$ factors occur frequently.  The naive
expectation would be that this is similarly true for non-toric bases,
although it would be important to initiate some systematic survey of
non-toric bases (perhaps, e.g., general hypersurfaces in toric
fourfolds), to confirm or contradict that hypothesis.
Such a survey would also give insight into whether the cycles needed
for remainder fluxes are indeed typical, as suggested in \cite{Braun:2014xka}.
For a given fourfold geometry, with multiple rigid gauge factors,
we
    can apply our construction to any rigid $E_7$ or $E_6$
    factor (while other gauge factors can serve as hidden
    sectors such as dark matter \cite{MorrisonTaylor4DClusters,Halverson:2016nfq}). We
    can then count the
    configurations of gauge-breaking flux explicitly, while
estimating the number of horizontal flux configurations
    using statistical methods \cite{DenefLesHouches}.
This can give a sense of the statistical likelihood of realizing the
Standard Model using the construction presented here for a given
geometry.  Combining these global and local analyses of large
classes of models in a systematic way could give a more precise
framework for characterizing the extent to which the construction
presented here is ``natural'' in the string landscape.
    
    \item We have focused here on the chiral part of the matter
    spectrum only, while the full matter spectrum also
    includes vector-like matter like the Higgs. Analyzing
    the
    vector-like spectrum requires explicit cohomology data
    from topologically nontrivial $C_3$ potential
    backgrounds.  These  are usually much harder to
    compute than $G_4$ flux, although recently
    analytical tools have been developed for some special
    cases of these
    \cite{Bies:2014sra,Bies:2021nje,Bies:2021xfh}; such analysis goes beyond the scope of this paper.
    On the other hand, we have a qualitative picture of the
    vector-like spectrum. Since we have started with a gauge
    group $G$ much larger than $\gsm$, generically there
    would be a large amount of vector-like matter, coming from the
    adjoint of $G$. It has been shown in
    \cite{BeasleyHeckmanVafaII} that it is
    impossible to remove all the vector-like exotics
    when the GUT group is $\SO(10)$ or higher\footnote{The argument
      given in \cite{BeasleyHeckmanVafaII} appears in a context where
 the gauge
      divisor is del Pezzo but the same argument holds whenever
the
 gauge divisor has an effective anti-canonical class, and has
      vanishing $h^{2,0}$.}, but
    it may be possible to remove the overly dangerous ones completely from the spectrum, such as $(\mathbf 3,\mathbf 2)_{-5/6}$ as demonstrated in \S\ref{sec:example}. We also expect the remaining
    vector-like matter to get large
    masses and lift from the low-energy theory. From this
    point of view, it has not been clear how the Higgs sector
    can be obtained with the right mass within  F-theory or any other
    approach for supersymmetric compactification of string
    theory. It is
    important to address this question if we want to fully
    realize the Standard Model in string theory.
    
    \item It is natural to consider $\U(1)$ extensions to our
    SM-like models, as extra $\U(1)$ factors can be easily
    constructed using the formalism of flux breaking. First,
    recall that some Cartan gauge bosons become massive due
    to vertical flux. In fact, they are still associated with
    global $\U(1)$ symmetries, although we expect that these
    symmetries are further (slightly) broken by other
    effects such as instantons \cite{Banks:2010zn}. Moreover, while $\U(1)$ gauge factors
    usually originate from a
    nontrivial Mordell-Weil group of rational sections in the global
    elliptic geometry
    \cite{AspinwallMorrisonNonsimply,MorrisonVafaII}, the
    $\U(1)$ factors from fluxes only depend on the local
    geometry on $\Sigma$, hence do not constrain the global
    geometry much. The resulting charges can easily be large,
    as shown in \S\ref{ssubsec:exotic}. Including these
    $\U(1)$'s in the models presented here can lead to extra selection
    rules and help resolve the puzzles in GUTs
    such as proton decay \cite{Marsano:2009wr,Grimm:2010ez},
    and is important in further studies of these SM-like
    models. In addition, it is interesting to explore the
    possibilities of large $\U(1)$ charges in 4D F-theory
    models from (vertical) flux breaking. (See e.g.
    \cite{Raghuram:2018hjn} for such an analysis in 6D F-theory
    models)
    
    \item Comparing with other tuned SM-like or GUT models,
    the origin of Yukawa couplings in our models is less
    clear. In the tuned models, only matter localized on
    curves $C$ is chiral and the Yukawa couplings are
    between
    three fields on $C$ ($CCC$), which are well understood by
    studying codimension-3 singularities (see, e.g.,
    \cite{WeigandTASI} for a review and further references). In contrast, chiral
    matter in our models may live on both the bulk of
    $\Sigma$ and on matter curves. Hence there are three
    possible types of Yukawa couplings: couplings between three
    fields on the bulk of $\Sigma$ ($\Sigma\Sigma\Sigma$),
    couplings between two fields on the matter curve $C$ and one
    field on $\Sigma$ ($CC\Sigma$), and the above $CCC$
    couplings \cite{BeasleyHeckmanVafaI}. It is
    natural to realize the Higgs on the bulk of $\Sigma$ since
    $\Sigma$ supports much vector-like matter, while a generic matter curve only supports chiral matter \cite{Bies:2020gvf}. The SM Yukawa couplings, which are between two
    chiral fields and the Higgs, thus should
correspond to $\Sigma\Sigma\Sigma$ and $CC\Sigma$ couplings.
    Nevertheless, rigid gauge groups %are 
can be
realized on
    $\Sigma$ with effective
    $-K_\Sigma$
(and therefore also
    $h^{2,0}(\Sigma)=0$), 
as in the explicit example of \S\ref{sec:example},
where $\Sigma\Sigma\Sigma$
    couplings are absent by the logic of
    \cite{BeasleyHeckmanVafaI}. Therefore to have
    the correct Yukawa couplings in this situation, extra tuning on
    fluxes must be done such that the chiral matter
    is localized on $C$ only.  While the tuning can
    be easily done in general, it is not possible in
    the example in \S\ref{sec:example} since
    $C_\mathbf{27}$ is trivial. Excluding
    this issue, we see no obstruction to having the Standard Model Yukawa
    couplings, but a rigorous construction is still lacking.
    There is a second issue specifically for $E_7$ models:
    as mentioned before,  codimension-3 singularities can arise in
    $E_7$ models with degrees $(4,6,12)$, which cannot be simply
    interpreted as $CCC$ couplings. This fact can also be seen from
    group theory, since $\mathbf{56}^3$ does not contain any
    singlets.
For a complete understanding of rigid $E_7$ flux breaking, the role of
fluxes through extra cycles associated with these singularities should
be better understood.
\end{itemize}

We hope to address some of these issues in future studies.

\acknowledgments{We would like to thank Lara Anderson, James
Gray, Patrick Jefferson, Manki Kim, Andrew Turner,  Yinan
Wang, and Timo Weigand for helpful discussions. 
We would also like to thank an anonymous JHEP referee for helpful
comments on an earlier version that led to improvements in the paper.
This work was supported by the
DOE under contract \#DE-SC00012567.}

\appendix

\section{Flux-induced St\"uckelberg mechanism}
\label{sec:Stuckelberg}

In this appendix, we review \cite{Donagi:2008ca,Grimm:2010ks} how vertical gauge-breaking flux
induces St\"uckelberg masses for the gauge bosons in broken
$\U(1)$ directions. As we will see, the masses are indeed
given by nonzero $\Theta_{i\alpha}$. For simplicity, here we
ignore all numerical factors and signs, which are not
important to the results.

This effect is perhaps most easily understood in the dual M-theory
picture. Consider M-theory compactified on $\hat Y$. We need
the parts of the supergravity action $S_{11D}$ involving
$G_4$:
\begin{equation}
    S_{11D}\supset\int_{\mathbb R^{2,1}\times\hat Y}\left(G_4\wedge *G_4+C_3\wedge G_4\wedge G_4\right)\,.
\end{equation}
Here the first term is the kinetic term for $G_4$ and the
second term is the Chern-Simons coupling. We now expand $C_3$
and $G_4$ with the following relevant terms:
\begin{equation}
    C_{3}\supset A^{\alpha}\wedge\left[D_{\alpha}\right]+A^{i}\wedge\left[D_{i}\right]\,,\quad G_{4}\supset F^{\alpha}\wedge\left[D_{\alpha}\right]+F^{i}\wedge\left[D_{i}\right]+G_{\rm int}\,,
\end{equation}
where $A$ are the $\U(1)$ gauge fields in 3D, $F=dA$, and
$G_{\rm int}$ is the flux in compactified or internal directions
i.e. the $G_4$ in the main text. We can then integrate over
$\hat Y$ and get the 3D effective action. Recall that
in the F-theory limit, $A^{\alpha}$ lives in chiral
multiplets and can be dualized into axions, while $A^{i}$
lives in vector multiplets giving the gauge bosons
in 4D. In particular, $A^{\alpha}$ and $A^i$
decouple in the kinetic term. The terms involving
$A^{\alpha}$ and its derivative are therefore
\begin{equation}
    S_{3D}\supset\int_{\mathbb{R}^{2,1}}\left(K_{\alpha
      \beta}F^{\alpha}\wedge*F^{\beta}+\Theta_{i\alpha}A^{i}\wedge F^{\alpha}\right)\,,
\end{equation}
where
\begin{equation}
    K_{IJ}=\int_{\hat Y}\left[D_{I}\right]\wedge*\left[D_{J}\right]\,,
\end{equation}
is the metric. There are also terms proportional to
$A^{\alpha}\wedge F^{i}$, but they are the same as
$A^{i}\wedge F^{\alpha}$ by integration by parts. 

We then construct the axion dual. First notice that since $dF^{\alpha}=0$, we can add a Lagrange multiplier $a_{\alpha}$ to the action i.e. a term $a_{\alpha}dF^{\alpha}$. Performing integration by parts, we have
\begin{equation}
    S_{3D}\supset\int_{\mathbb{R}^{2,1}}\left(K_{\alpha \beta}F^{\alpha}\wedge*F^{\beta}+\left(da_{\alpha}+\Theta_{i\alpha}A^{i}\right)\wedge F^{\alpha}\right)\,.
\end{equation}
The equation of motion gives
\begin{equation}
    *F^{\beta}=K^{\alpha \beta}\left(da_{\alpha}+\Theta_{i\alpha}A^{i}\right)\,.
\end{equation}
We finally integrate out $F^\alpha$ and get
\begin{equation}
    S_{3D}\supset\int_{\mathbb{R}^{2,1}}\left(K^{\alpha\beta}\left(da_{\alpha}+\Theta_{i\alpha}A^{i}\right)\wedge*\left(da_{\beta}+\Theta_{j\beta}A^{j}\right)\right)\,.
\end{equation}
By gauge transformations, the gauge fields $A^i$ can ``eat''
the axions $a_\alpha$ and become massive as long as there are
enough axion fields. The masses are determined by the
eigenvalues of the mass matrix
$K^{\alpha\beta}\Theta_{i\alpha}\Theta_{j\beta}$. Its
null space, hence massless $\U(1)$ directions, corresponds to
linear relations between $\Theta_{i\alpha}$ i.e.
$c_i\Theta_{i\alpha}=0$ for all $\alpha$. This justifies the flux constraints
imposed in the main text for gauge breaking.

\section{Embeddings of Standard Model gauge group into $E_7$}
\label{sec:embeddingcount}
In this Appendix, we count different embeddings of $\gsm$
into $E_7$ giving SM matter representations. It is stated in
\S\ref{ssubsec:SMchiral} that the root embedding of
$\SU(3)\times\SU(2)$ is unique up to automorphisms, and there
are 4 distinct choices of hypercharge $\U(1)$. Here we prove
these claims.

To prove the uniqueness of the root embedding  up to automorphisms
we proceed in a somewhat
explicit constructive fashion.
We can describe $E_8$ explicitly as a lattice consisting of all points
\begin{equation}
  (x_1, x_2, \ldots, x_8) \in \{(\Z)^8\cup (\Z +1/2)^8: \sum_i x_i
  \equiv 0 \;({\rm mod}\ 2)\}\,.  
\label{eq:}
\end{equation}
The roots of $E_8$ are the 240 elements of this lattice satisfying $r
\cdot r = 2$.
$E_7$ can be realized as the
orthogonal complement of any root $r_8$ of $E_8$, so without loss of
generality we pick $r_8 =(1, -1, 1, -1, 1, -1, 1, -1)/2$.
There are 126 roots $r$ of $E_8$ satisfying $r \cdot r_8 = 0$,
corresponding to the roots of $E_7$.  To embed $SU(3) \subset E_7$ as
a root embedding, we wish to choose roots $r_1, r_2 \in E_7$ such that
$r_1 \cdot r_2= -1$.  Choosing arbitrarily $r_1 = (1, 1, 0, 0, 0, 0,
0, 0)$ from the 126 equivalent roots of $E_7$, there are 32 roots
satisfying the condition on $r_2$, from which we pick arbitrarily $r_2
= (0, -1, -1, 0, 0, 0, 0, 0)$.  There are 30 roots of $E_7$ that are
perpendicular to $r_1, r_2$, so we embed the $SU(2)$ with the
arbitrary choice $r_7 = (0, 0, 0, 1, 1, 0, 0, 0)$.  Assuming
momentarily that all 32 choices of $r_2$ give 30 choices of $r_7$
(which we will prove below shortly) this gives $126*32*30 =120,960$
root embeddings of $SU(3) \times SU(2)$ into $E_7$.

We now show that
our given choice is equivalent to the one illustrated in
Figure~\ref{dynkine7}.  To identify the root associated with node 3 in
that diagram, we need a root $r_3 \in E_7$ such that $r_3 \cdot  r_1 =
0, r_3 \cdot r_2 = -1, r_3 \cdot r_7 = -1$.  There are 4 such roots,
among them we pick $r_3 =(0, 0, 1, -1, 0, 0, 0, 0)$.  Continuing in
this fashion, there are 3 choices for $r_4$, and 2 choices for $r_5$,
after which $r_6$ is uniquely determined.  Multiplying out
$120,960*24*240 = 696,729,600$, which is exactly the size of the
automorphism group of $E_8$.  Given one choice of embedding given by
the sequence of roots described above, each automorphism of $E_8$ will
give a distinct embedding.  Thus, we must have at least this many
independent sequences of choices based on the equivalent embeddings.
If there were any inequivalent root embeddings, they would have given
rise to a larger number of choices at some step in the process.  This
proves that indeed all of the root embeddings are  equivalent at each stage.
Note that there are also exotic non-root embeddings of $SU(3) \times
SU(2)$ into $E_7$\footnote{We would like to thank Andrew Turner for
  discussions on this point.}, but these cannot be realized by flux breaking and
are not relevant to the discussion here.

We can now relate this analysis to the choices of hypercharge.
Without loss of generality, we can now put the non-abelian
part of $\gsm$ in nodes $1,2,7$ in Figure \ref{dynkine7}. Let
the hypercharge of $R'$ with given $b_{i'}$ be $q_Y=a_{i'} b_{i'}$, where $i'=3,4,5,6$ and $a_{i'}$ are numbers to be
solved. We then break the $\mathbf{56}$ into representations of
$\gsm$ and require them to be the SM representations. There
is only one $(\mathbf 3,\mathbf 2)$, which has
$b_{i'}=(2,3/2,1,1/2)$. Therefore,
\begin{equation}
    2a_3+3a_4/2+a_5+a_6/2=1/6\,.
\end{equation}
Similarly, by looking at $(\bar{\mathbf 3},\mathbf 1)$ and $(\mathbf 1,\mathbf 2)$, we get
\begin{align}
    a_3+3a_4/2+a_5+a_6/2&=-2/3\;\mathrm{or}\;1/3\,,\\
    a_3+a_4/2+a_5+a_6/2&=-2/3\;\mathrm{or}\;1/3\,,\\
    a_4/2+a_5+a_6/2&=\pm 1/2\,,\\
    a_4/2+a_6/2&=\pm 1/2\,,\\
    a_4/2-a_6/2&=\pm 1/2\,.
\end{align}
It is then straightforward to deduce that the only
possibilities of $a_{i'}$ are
\begin{equation}
    a_{i'}=(5/6,-1,0,0),(-1/6,1,-1,0),(-1/6,0,1,-1),(-1/6,0,0,1)\,.
\end{equation}
These four $a_{i'}$ correspond to the four choices in Eq.\ (\ref{hyperchargeChoice}), and the four roots that enhance
$\SU(3)\times\SU(2)$ to $\SU(5)$, which are equivalent under
automorphisms as described above. This finishes our proof.

Note that we have assumed here that the U(1) hypercharge assignment
for the states coming from the $\mathbf{56}$
gives the SM values, with no exotics.  This does not rule
out a choice of U(1) where the states coming from the $\mathbf{56}$
include some exotics and omit some SM states, while the
states coming from the $\mathbf{133}$ can complete the SM
states and contain other exotics, as we found explicitly for the
breaking pattern described in \S\ref{ssubsec:exotic}.  While in that
case, there was no flux choice that gives just the SM
chiral matter content with no exotics, we have not ruled out the
possibility that some other U(1) hypercharge assignment may allow in
principle for a similar situation, where fine tuning of the fluxes may
reduce to only SM chiral matter.  We leave a more detailed
investigation of this question for further work.

\section{Resolution of $E_6$ model}
\label{sec:resolution}

In this appendix, we describe more about the resolution of
$E_6$ model used in the main text. As a starting point, a
generic $E_6$ model can be described by a Tate model \cite{BershadskyEtAlSingularities,KatzEtAlTate}, where
$Y$ is given by the locus of
\begin{equation}
    y^2+a_{1,1}sxyz+a_{3,2}s^2yz^3=x^3+a_{2,2}s^2x^2z^2+a_{4,3}s^3xz^4+a_{6,5}s^5z^6\,,
\end{equation}
where the class of $a_{i,j}$ is $-(iK_B+j\Sigma)$. We now
resolve this model by performing blowups. We denote
\begin{equation}
    Y_1\stackrel{(x,y,s|e_{1})}{\longrightarrow}Y\,,
\end{equation}
as the blowup from $Y$ to $Y_1$ by the redefinition
\begin{equation}
    x\rightarrow xe_1\,,\quad y\rightarrow ye_1\,,\quad s\rightarrow se_1\,.
\end{equation}
The resulting locus $e_1=0$ is a divisor in the ambient
space, denoted by $E_1$. Using the same notation, we can then
write down the resolution as the following steps \cite{Esole:2017kyr,Bhardwaj_2019,Jefferson:2021bid}:
\begin{equation}
    \hat Y\stackrel{(y,e_4|e_6)}{\longrightarrow}Y_5\stackrel{(y,e_3|e_5)}{\longrightarrow}Y_4\stackrel{(e_2,e_3|e_4)}{\longrightarrow}Y_3\stackrel{(x,e_2|e_3)}{\longrightarrow}Y_2\stackrel{(y,e_1|e_2)}{\longrightarrow}Y_1\stackrel{(x,y,s|e_1)}{\longrightarrow}Y\,.
\end{equation}
This resolution smooths out all singularities on $Y$ up to codimension-3. The exceptional divisors on $\hat Y$ are given by
\begin{align}
    D_1 &= E_5\cap \hat Y\,, \nonumber \\
    D_2 &= E_6\cap \hat Y\,, \nonumber \\
    D_3 &= (-E_1+2E_2-E_3-E_4)\cap \hat Y\,, \nonumber \\
    D_4 &= (E_1-2E_2+E_3+2E_4-E_6)\cap \hat Y\,, \nonumber \\
    D_5 &= (E_3-E_4-E_5)\cap \hat Y\,, \nonumber \\
    D_6 &= (E_1-E_2)\cap \hat Y\,.
\end{align}
Using the above information, the intersection numbers between
divisors on $\hat Y$ can then be computed using the
techniques in \cite{Esole:2017kyr}.

\bibliography{references}
\bibliographystyle{JHEP}

\end{document}